\documentclass[12pt]{iopart}

\usepackage{iopams}  
\usepackage{graphicx}

\begin{document}

\title[Potential and Feshbach $s$-wave resonances]{Potential and Feshbach $s$-wave resonances in coupled atomic collision channels}

\author{G. Andrade-S\'anchez and V. Romero-Roch\'in}

\address{Instituto de F\'isica, Universidad Nacional Aut\'onoma de M\'exico \\
Apartado Postal 20-364, 01000 Cd. M\'exico, Mexico}

\ead{romero@fisica.unam.mx}
\vspace{10pt}
\begin{indented}
\item[]May 2023
\item[]
\end{indented}

\begin{abstract}
We discuss $s$-wave scattering in an atomic binary collision with two coupled channels, tunable by an external magnetic field, one channel open and the other closed for the incident energies considered.  The analysis is performed with a stylized model of square-well potentials. This simplification allows for a pedagogically thorough discussion of the different scattering resonances that appear in coupled channels. One of the them, the potential resonances at vanishing energy, occur as a bound state of the coupled system emerges, in turned tuned at a very precise value of the external field. The other resonances, described by Feshbach theory, occur when the incident energy is near a bound state of the closed channel, as if it were decoupled from the open channel. These resonances exist for values of the external field above a particular threshold value. Besides the potential intrinsic value of this study in a quantum mechanics course, as the analysis can be performed with minor numerical calculations, it is also an aid for the understanding of current research advances in the exciting field of ultracold gases. 
\end{abstract}

\section{Introduction}

Experiments in ultracold Bose\cite{Ketterle,Cornell,Hulet} and Fermi\cite{Jin,Kett-Zwier,Hulet-rev} gases, with its unprecedented external control of atomic interactions, opened a new window into novel quantum macroscopic behaviors, generally called the Bose-Einstein Condensation\cite{Pethick,Dalfovo,Bloch} for the Bose gases and the BEC-BCS crossover\cite{Legget,Giorgini,Ohashi} for the Fermi ones. This is a smooth transition between two superfluid phases, in one extreme a gas of fermionic Cooper pairs in the Bardeen-Cooper-Schriefer (BCS) state and, in the other, a Bose-Einstein condensate of bosonic paired atoms. This crossover has allowed us to gain a better understanding of superfluidity\cite{nobel}, generation of exotic topological phases\cite{Tiurev}, high-temperature superconductivity\cite{Muller,Randeria,Tsuei}  and the structure of neutron stars\cite{Chamel}, among others \cite{Strinati}. While these phenomena emerge due to thermodynamic and many-body effects, the core of accessing them rests on the experimental ability of externally controlling the pairwise atomic interactions or, in other words, the binary atomic collisions. This control is achieved by exploiting the fact that alkaline atoms, such as $^6$Li, $^{23}$Na, $^{40}$K and $^{87}$Rb, typically used in ultracold experiments, have hyperfine energy levels that can be externally tuned by magnetic fields by means of the Zeeman effect\cite{Verhaar,Moerdijk,Houbiers,OHara,Lange,Zurn,Julienne}. This leads to collisions between atoms with several internal states or, as also known in the literature, as collisions with multiple channels.\\

One outstanding feature that emerges with the control of the Zeeman splittings of those hyperfine levels, by means of an external magnetic field, is the tuning of the interatomic interactions, resulting in being either effectively repulsive or effectively attractive, the former essential for bosonic and the later for fermionic superfluidity. This drastic and contrasting change of the character of the interactions is tied to the occurrence of a scattering resonance \cite{Verhaar,Moerdijk,Houbiers}, commonly called a ``Feshbach resonance'', after the seminal studies by Feshbach on multiple channel scattering resonances \cite{Fesh1,Fesh2,Fesh3}. This quantum control gave way to the spectacular advances that we have seen in the field of ultracold gases. Detailed accounts of experimental determinations of these Feshbach resonances can be found in  \cite{OHara,Lange,Zurn,Julienne} and in the excellent reviews  \cite{Timmermans,Duine,Chin}.\\

And here is the main goal of this paper, namely, to present an explicit, thorough and pedagogical calculation of the scattering properties of two atoms with two internal channels only, using the simplest model of atomic interactions, the so-called ``square-well'' potentials. On the one hand, in essentially all quantum mechanics textbooks, e.g. \cite{LLQM,Liboff,Cohen}, scattering is presented for single-channel cases, that is,  one interatomic potential only, and rarely the resonance phenomenon is addressed. And, on the other hand, in most of the specialized articles dealing with this subject, e.g. \cite{Verhaar,Moerdijk,Houbiers,OHara,Lange,Zurn,Julienne,Timmermans,Duine,Chin}, the reader is expected to fill the gaps. For sure, there are authoritative monographies on scattering theory, such as the books by Newton \cite{Newton} and Taylor \cite{Taylor}, that besides being quite intimidating even for professional researchers, do not explicitly approach the atomic problem of current interest, having been written for studying nuclear collisions mostly. Certainly, many aspects of the present paper have their detailed foundation on those writings and will constantly be referred throughout.  
There are some articles that deal with models similar to the one here discussed but, being articles written for specialists\cite{Timmermans,Duine,Chin,Kokkelmans,Gurarie,Wasak}, do not fully give the needed details. We must also mention two articles that we are aware of that deal with the subject of this article in a pedagogical manner as well, one is an old paper by Fraser and Burley \cite{Fraser}, published in this journal in 1982,  and the other, much more recently, by Taron \cite{Taron}. Ours complements those with a more detailed and different approach. In addition, as a minor clarifying aspect of the present article, and as we will repeatedly emphasize, there appears to be a technically incorrect referral to the mentioned ``Feshbach resonances'' that, in practical terms, do not affect the understanding of the physics underneath: we shall argue that those resonances are actually ``potential resonances'' and the ones studied by Feshbach are of a different nature. 
Thus, reiterating, the purpose of this article is the presentation and discussion of the scattering properties of atomic collisions with coupled internal states, or channels, as well as the involved resonances. The presentation should be appropriate for undergraduate and graduate students, with knowledge of quantum mechanics and the basics of scattering theory.\\

\section{Scattering in two channels}

As mentioned above, textbook discussions of scattering in three dimensions consider a potential with spherical symmetry, $V(r)$, that reduces the problem to the relative coordinate ${\bf r} = {\bf r}_1 - {\bf r}_2$ of two particles\cite{LLQM,Liboff,Cohen}. This can represent a true collision of two atoms by invoking the Born-Oppenheimer approximation and considering that the potential $V(r)$ corresponds to the electronic ground state of the binary molecule thus formed, although the scattering is for positive incident energies. Atoms, for sure, are much more complicated than that and, in particular, their fine and hyperfine structure can play a significant role. Such is the case of alkaline atoms whose ground state is typically degenerate in the hyperfine levels of the nuclear-electronic spin interactions \cite{Verhaar,Moerdijk,Houbiers}. Dealing with this full problem is a formidable task since the number of such levels can be quite large due to the values of the nuclear spin; for instance if each atom has nuclear spin $I = 1$ and electronic spin $s = 1/2$, the number of hyperfine levels of the binary system is twelve. The presence of an external magnetic field, by means of Zeeman interaction, can split those levels and, since the obtained energy separation is very small for atomic scales, say in the megahertz, the resulting collision involves then several potentials $V_\alpha(r)$, one for each level and, very importantly, during the collision process those potentials are coupled. It is this small energy splitting the responsible for the ``breaking'' of the Born-Oppenheimer approximation between the hyperfine levels and, thus, the consideration of their coupling is necessary. The collision is then said to occur in several channels, with some of them ``open'' and other ``closed'', depending on the incident energy of the atoms. Indeed, researchers in this field perform sophisticated and accurate calculations of these situations that experimentalists use for the assessment and analysis of their data\cite{Verhaar,Moerdijk,Houbiers,OHara,Lange,Zurn,Julienne}. However, and experiments have shown this can be the case, in many cases the collision involves mainly only two hyperfine levels. This fortunate situation not only reduces the mathematical difficulties, but also allows for a simpler understanding of the underlying physics, in particular, the collision resonances that thus arise. \\

\begin{figure}[htbp]
\begin{center}
\includegraphics[width=0.5\linewidth]{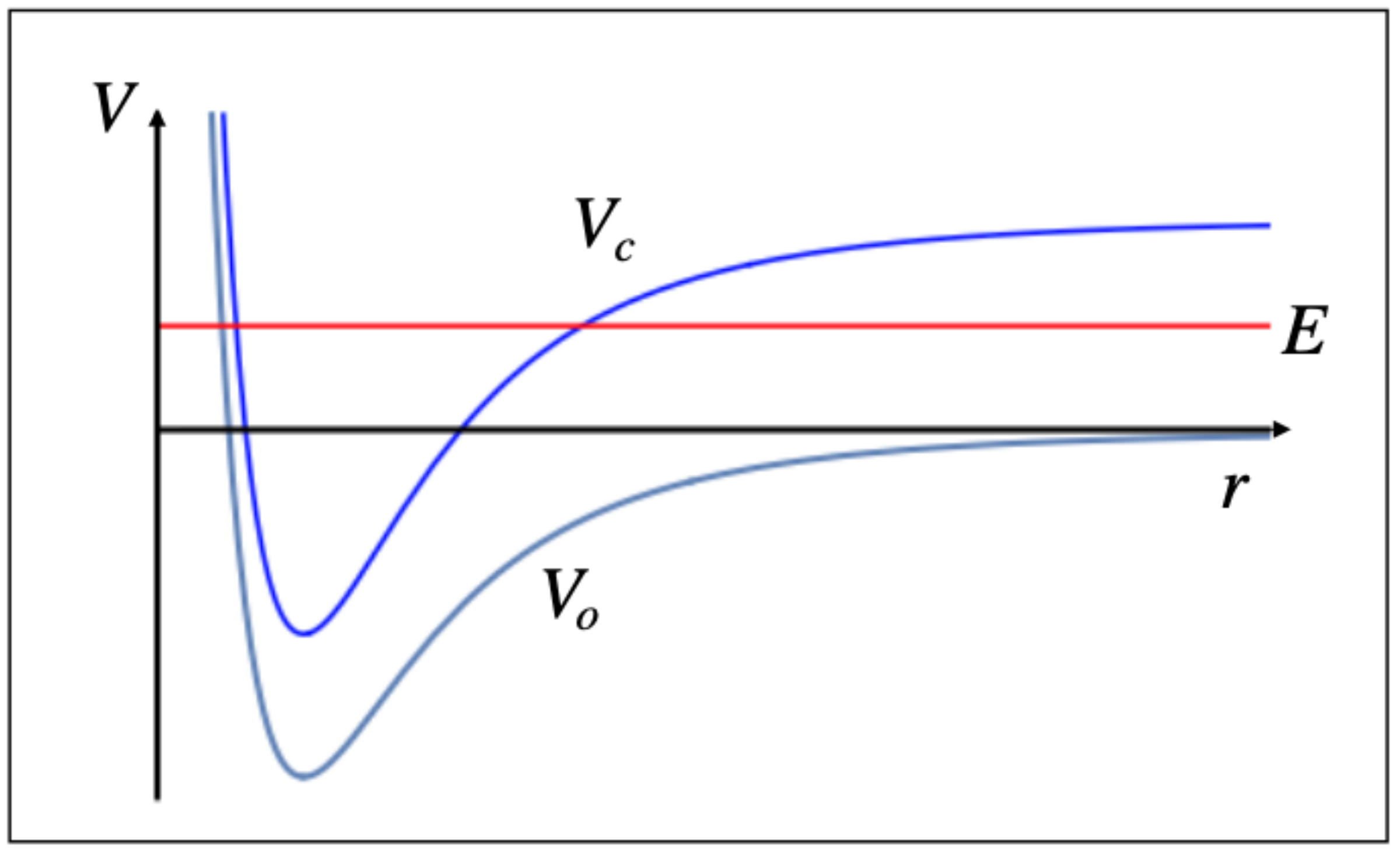}
\end{center}
\caption{(Color online) Schematic representation of two-channel potentials, $V_c(r)$ is the closed channel potential and $V_o(r)$ the closed one. For large $r$ the energy separation of the potential is given by $\Delta \mu {\cal B}$. For short distances the channels are coupled by the off-diagonal term $V_{hf}(r)$ in the Hamiltonian (\ref{H}). The scattering process we consider is for energies $E$ between the two channels.
} \label{Fig1}
\end{figure}

The above simplification to two {\it coupled} hyperfine internal states, or scattering {\it channels}, can be cast with the following Schr\"odinger equation for this problem, in the relative coordinate ${\bf r} = {\bf r}_1 - {\bf r}_2$, 
\begin{equation}
\fl \left(\begin{array}{cc}
-\frac{\hbar^2}{2m} \nabla^2 + V_o(r) & V_{hf}(r) \\
V_{hf}(r) & -\frac{\hbar^2}{2m} \nabla^2 + V_c(r) + \Delta \mu {\cal B}
\end{array}\right)
\left(\begin{array}{c}
\psi(\vec r) \\
\phi(\vec r)\end{array}\right) = E \left(\begin{array}{c}
\psi(\vec r) \\
\phi(\vec r)\end{array}\right) \>.\label{H}
\end{equation}
Here, $V_c(r)$ and $V_o(r)$ are the interatomic potential of the two channels in the absence of coupling between them, see figure \ref{Fig1}, with potential ranges $R_c$ and $R_o$, and vanishing in the asymptotic region $r \gg R_c, R_o$, $V_c(r) \to 0$ and $V_o(r) \to 0$. The label $o$ stands for {\it open} and the label $c$ for {\it closed}. These labels imply that the channels are decoupled in the asymptotic region with a Zeeman-like separation $\Delta \mu {\cal B}$, a notation inherited from the mentioned ultracold gases experiments, with $\Delta \mu$ the effective magnetic dipole moment of the hyperfine states and ${\cal B}$ an external magnetic field. The coupling $V_{hf}(r)$, where  $hf$ stands for {\it hyperfine}, is assumed to be different from zero within the interatomic potential ranges, namely, $R_{hf} \sim R_c, R_o$. Although for explicit atomic calculations it is important to deal with exchange issues, depending if the atoms are fermions or bosons, we will ignore them for simplicity. \\

The scattering situation we want to describe is for incident energies $E$ such that, in the scattering or asymptotic region, their values are between the open and the closed potential channels, namely $0 < E < \Delta \mu {\cal B}$, thus justifying their labels; see figure (\ref{Fig1}). Furthermore, we shall study $s$-wave scattering only, corresponding to zero angular momentum $l = 0$, since this is the dominant contribution for low energy, usual in cold collisions.  In this approximation the wavefunctions are spherically symmetric and Schr\"odinger equation can be further simplified as,
\begin{equation}
\fl \left(\begin{array}{cc}
-\frac{\hbar^2}{2m} \frac{d^2}{dr^2} + V_o(r) & V_{hf}(r) \\
V_{hf}(r) & -\frac{\hbar^2}{2m} \frac{d^2}{dr^2} + V_c(r) + \Delta \mu {\cal B}
\end{array}\right)
\left(\begin{array}{c}
u(r) \\
v(r)\end{array}\right) = E \left(\begin{array}{c}
u(r) \\
v(r)\end{array}\right) \>,\label{Hcoupled}
\end{equation}
where $v(r) = r \psi(r)$ and $u(r) = r \phi(r)$. 

The above considerations indicate that for $0 < E < \Delta \mu {\cal B}$, in the scattering region, the solution must be of the form
\begin{equation}
\left(\begin{array}{c}
u(r) \\
v(r)\end{array}\right) \approx 
\left(\begin{array}{c}
A(E) e^{ikr} + B(E) e^{-ikr} \\
C(E) e^{-\kappa_B r}\end{array}\right)
\>\>\>\>\textrm{for}\>\>r \gg R_c, R_o, R_{hf}\> ,\label{asymp}
\end{equation}
where  $k = \sqrt{2mE/\hbar^2}$ and $\kappa_B = \sqrt{2m(\Delta \mu B-E)/\hbar^2}$. This expression tells us that scattering occurs in one channel only, the open one, while in the closed one the solution vanishes for large separation distances. Therefore, following the usual treatment of scattering within a single potential, the scattering solution in the open channel can be cast as, \cite{LLQM,Liboff,Cohen}
\begin{equation}
u(r) \approx B(E)  \left( e^{-ikr} - e^{2i \delta_0} \>e^{ikr} \right) \>\>\>\>\textrm{for}\>\>r \gg R_c, R_o, R_{hf}\>,\label{uscat}
\end{equation}
with $\delta_0$ the $s$-wave phase shift. Comparison with solution (\ref{asymp}) yields the so-called $S$-matrix of the scattering problem, 
\begin{equation}
e^{2i \delta_0} = - \frac{A(E)}{B(E)} \>\>\>\>\>\textrm{for}\> 0 < E < \Delta \mu {\cal B} \>.\label{phashift}
\end{equation}
As we learn from any basic analysis of binary collisions, all the scattering properties can be found from the knowledge of the phase shifts of the scattered wave function and, since we are limiting to the $s$-wave approximation, in the present case those properties can be found via the knowledge of the phase shift $\delta_0(E)$ or, for that matter, from the $S$-matrix (\ref{phashift}). However, in order to find this quantity, or the coefficients $A(E)$ and $B(E)$, the asymptotic solution (\ref{asymp}) must be matched at short separation distances with the full solution of the coupled Hamiltonian given by (\ref{Hcoupled}). This will be done explicitly with a stylized model further below. A very important aspect to keep in mind is the fact that for $r \lesssim R_c, R_o, R_{hf}$ there is no longer a distinction of which channel is closed and which one is open. Of course, if the coupling $V_{hf}(r)$ is weak, there will be remnants of which potential plays one role or the other, but not in general.\\

Knowledge of the phase shift allows to find, on the one hand, the $s$-wave scattering cross section,
\begin{equation}
\sigma_0(E) = \frac{4 \pi}{k^2} \sin^2 \delta_0(E) \>,\label{sigma0}
\end{equation}
which is a function of the incident energy $E$. On the other hand, as
$s$-wave scattering becomes prominent for low energy collisions, $E \to 0^+$, an important quantity that characterizes the process is the scattering length $a$, given by,
\begin{equation}
 a = - \lim_{k \to 0^+} \frac{1}{k} \tan \delta_0 \>.\label{a0}
 \end{equation}
While $a$ does not depend on the energy anymore, it still does depend on the potential parameters. The scattering length $a$ of binary collisions plays a profound role on the understanding of the many-body properties of quantum Fermi and Bose gases \cite{Pethick,Dalfovo,Bloch,Timmermans,Duine,Chin}. Notice that if the phase-shift is small, $\delta_0 \ll 1$, then $\delta_0 \approx - k a$, a common way to introduce the scattering length; however, this is not always true, specially in the vicinity of a potential resonance, as discussed further below.\\

In the following sections we will analyze in detail the scattering problem of the two-channel binary collision formulated above with a stylized model of ``square-well'' potentials. We will focus on the elucidation of the behavior of the phase shift $\delta_0$ or, equivalently, the $S$-matrix given by (\ref{phashift}). The main goal of such an analysis is the understanding of the scattering resonances, identified as the energies $E$ where the cross section becomes a maximum, namely, when the scattering strong. Although there is not a ``formal'' definition of what it is considered a resonance in a collision, one can identify them as ``peaks'' in the cross section $\sigma_0$, as a function of incident energy $E \ge 0$\cite{Taylor}. The origin of these peaks can be stated a bit loosely here, and shown in the text,  as when the incident energy is close to {\it resonance} with a bound state, or to a remnant of it. More technically, observing expression (\ref{sigma0}) for $\sigma_0$, one finds that a peak, at a given $E$, cannot exceed the envelope $4\pi/k^2$, although it can ``touch'' it. Therefore, for purposes of the present discussion we can consider a collision resonance at, say, $E=E_{\rm res}$, when the peak yields $\sigma_0(E_{\rm res}) \approx 4\pi/k_{\rm res}^2$ with $k_{\rm res} = \sqrt{2m E_{\rm res}/\hbar^2}$, which in turn implyies that the phase shift is approximately $\delta_0 \approx \pi/2$ at the resonance. As we will explicitly show, the appearance and location of the resonances is closely tied to the existence of poles of the $S$-matrix, when this quantity is analytically continued to the complex $E$-plane. As seen from the expression for the $S$-matrix (\ref{phashift}), its poles are found when $B(E) = 0$. Notice also from the asymptotic solution (\ref{asymp}) that the bound states, with $E < 0$ real, require that in order to obtain a convergent solution, it must be true that $B(E) = 0$ as well, since $k = i \sqrt{2m |E|/\hbar^2}$ is then purely imaginary. This gives the additional formal result that the poles of the $S$-matrix, for real negative energies $E < 0$, are the eigenenergies of the bound states. For $E$ complex, with ${\rm Re} \> E > 0$ and ${\rm Im} \> E < 0$, the poles indicate resonances, such as those described by Feshbach theory, as will be discussed in this article.\\

For the two-channel collision we shall find that there exist two types of different resonances, one that we shall call {\it potential} resonances, already present in one-channel scattering \cite{LLQM,Liboff,Cohen} and others that are those predicted by the theory of Feshbach\cite{Fesh1,Fesh2,Fesh3}. The former are associated with the appearance of bound states of the {\it fully coupled} system, as its physical parameters are varied, say, as the field ${\cal B}$ is changed; and, very importantly, the emergence of the potential resonances are concomitant with the divergence of the scattering length. The latter resonances, those studied by Feshbach among others, are a consequence of the presence of the two channels, one closed and the other open. We will see that these  resonances appear very near the energy of a bound state of the closed channel, as if it were {\it decoupled} from the open one. 
It is important to stress that these states are not true bound states of the coupled system but their presence, nevertheless, affect the behavior of the cross section as if the full system were bound by them for a brief period of time \cite{Taylor}. As already mentioned, we shall show that 
the potential resonances have been commonly identified as ``Feshbach resonances'' in the literature and that the resonances predicted by the theory of Feshbach are of a different nature, namely, of the second case just described.\\

In Section 3 we will study the problem of the potential resonance using the square-well potentials model. Section 4 is devoted to an analysis of the poles of the $S$-matrix and their relationship to the resonances. In Section 5 we will first revisit the derivation of the resonances studied by Feshbach, then we will make a comparison of such an approximated prediction with exactly calculated cross sections of the model at hand, thus assessing its range of validity. We shall then see that these resonances are of a different nature than those of Section 3.

\section{Potential resonances in two coupled channels}

\begin{figure}[htbp]
\begin{center}
\includegraphics[scale=0.3]{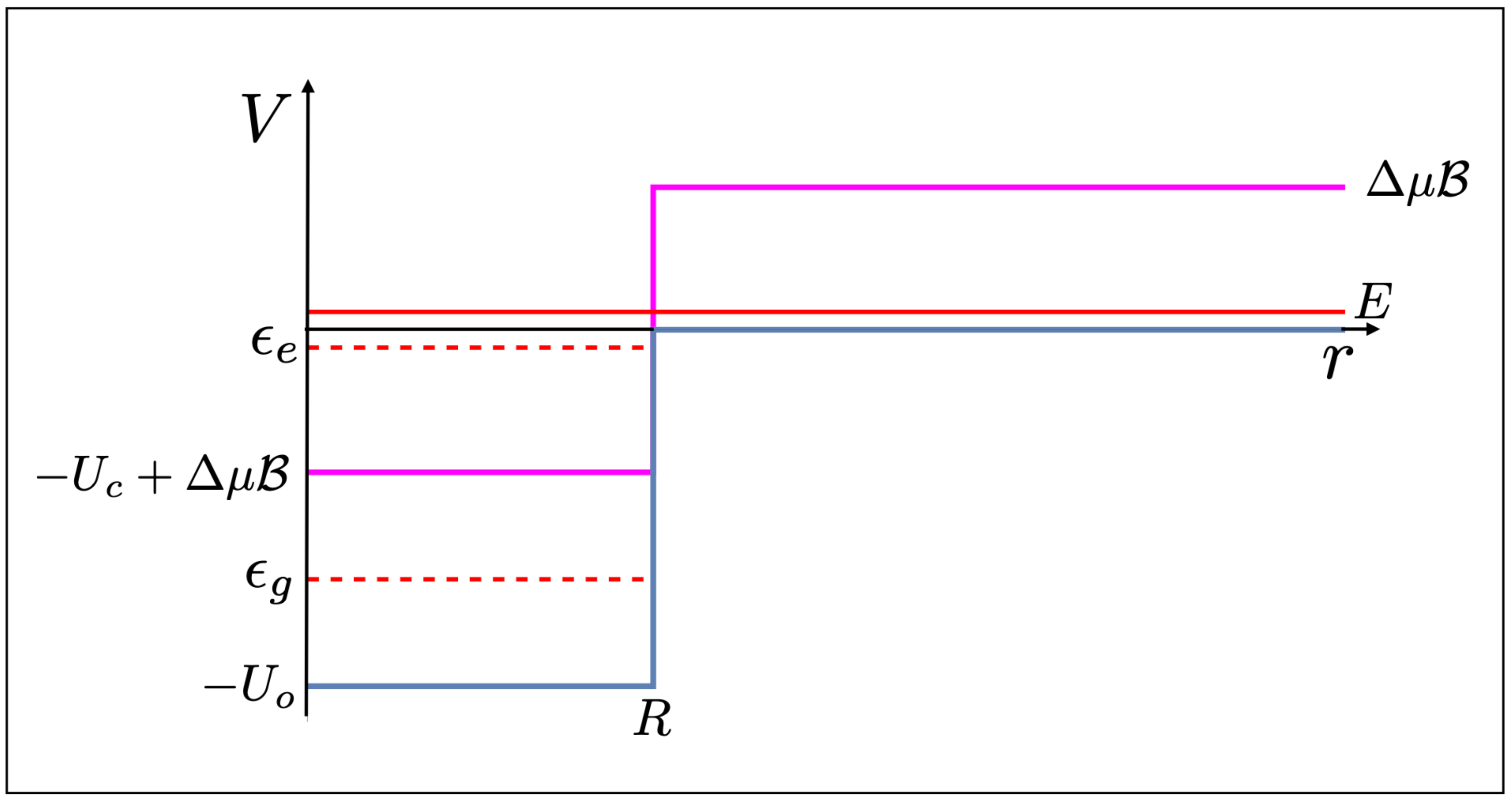}
\end{center}
\caption{(Color online) Schematic illustration of the two-channel model. $-U_o$ and $-U_c + \Delta \mu {\cal B}$ are the open (blue) and shifted closed (magenta) channels. As described in the text, with no loss of generality, we consider cases where the full coupled two-channel system allows for the existence of two bound states $\epsilon_g$ and $\epsilon_e$ only. Due to the coupling $U_{hf}$, not shown in the figure, as the external magnetic field ${\cal B}$ is increased, the excited bound state energy $\epsilon_e$ shifts ``upwards'' until it vanishes at  a threshold value ${\cal B}^\star$, with the emergence of a resonance potential at arbitrarily small scattering energies, $E \to 0^+$. The figure also pretends to illustrate this resonance condition, with ${\cal B} \lesssim {\cal B}^\star$.
} \label{esquema}
\end{figure}

We now return to the study of the two coupled channels, described by Schr\"odinger equation (\ref{Hcoupled}), introducing a simple model in order to be able to explicitly calculate general aspects of the scattering process. For this, we consider the atomic potentials to be square-well potentials and, for simplicity as well, the hyperfine coupling is also taken to be a constant, namely,
\begin{equation}
V_c(r), V_o(r), V_{hf}(r) = \left\{
\begin{array}{ccc}
-U_c, - U_o, U_{hf} & \textrm{if} & r \le R \\
 & & \\
 0 &  \textrm{if} & r > R \end{array}\right.\>. \label{potentials}
\end{equation}
Note that all potentials strengths are positive, $U_c > 0, U_o > 0, U_{hf} > 0$, and 
for the sake of explicit calculations we have assumed their respective ranges to be the same, $R_c = R_o = R_{hf}= R$. Figure \ref{esquema} illustrates this model, as well as the scattering and bound states we study here. Evidently, the structure of the bound states and the scattering properties depend on all these parameters and on the external magnetic field ${\cal B}$. With no loss of generality and, inspired by the way ultracold current experiments are conducted, we shall keep all potential parameters fixed and consider the variation of the external magnetic field as the only tuning quantity. This will result in a simple shift in energy of the open channel potential, $-U_c + \Delta \mu {\cal B}$ for $r < R$, and $\Delta \mu {\cal B}$ for $r > R$.\\

With this model at hand one can analytically solve Schr\"odinger equation (\ref{Hcoupled}) for $0 < E < \Delta \mu {\cal B}$, some details given in Appendix A, and one can find the important coefficient $B(E)$ of the $r > R$ solution (\ref{asymp}), yielding
\begin{eqnarray}
\fl B(E) & =&  {\cal C}_0 e^{ikR} \left(\cos^2 \frac{\theta}{2} \left(k_- \cos k_- R - i k \sin k_-R\right) \left(k_+ \cos k_+R + \kappa_B \sin k_+R\right) + \right. \nonumber \\
\fl &&\left. \sin^2 \frac{\theta}{2}  \left(k_+ \cos k_+ R - i k \sin k_+R\right) \left(k_- \cos k_-R + \kappa_B \sin k_-R\right) \right)\>,\label{BEpos}
\end{eqnarray} 
where ${\cal C}_0$ is a real normalization constant and,
\begin{eqnarray}
k & = & \sqrt{\frac{2mE}{\hbar^2}} \\
\kappa_B & = & \sqrt{\frac{2m(\Delta \mu {\cal B}-E)}{\hbar^2}} \\
k_\pm & = & \sqrt{\frac{2m(\epsilon_{\pm} + E)}{\hbar^2}} \>,\label{ks}
\end{eqnarray}
with
\begin{equation}
\epsilon_{\pm} = \frac{U_o + U_c - \Delta \mu {\cal B}}{2} \pm \sqrt{\left(\frac{U_o - U_c + \Delta \mu {\cal B}}{2}\right)^2 + U_{hf}^2} \>, \label{epm}
\end{equation}
and 
\begin{eqnarray}
\cos \theta &=&  \frac{U_o - U_c + \Delta \mu {\cal B}}{2\sqrt{\left(\frac{U_o - U_c + \Delta \mu{\cal B}}{2}\right)^2 + U_{hf}^2}} \nonumber \\
\sin \theta & = & \frac{U_{hf}}{\sqrt{\left(\frac{U_o - U_c + \Delta \mu {\cal B}}{2}\right)^2 + U_{hf}^2}} \>.\label{teta}
\end{eqnarray}
The weak coupling limit is when $U_{hf}$ is very small, $\theta \to 0$, and the strong one when $U_{hf}$ is very large, $\theta \to \pi/2$. In the extreme case $\theta = \pi/2$, the present problem reduces to the problem of a single open channel. \\

We recall that the $B(E)$ coefficient is the denominator of the $S$-matrix (\ref{phashift}). In the specialized literature this coefficient is known as the Jost function \cite{Newton,Taylor}. Notice that only for $E > 0$ the phase shift is real and, therefore, we can choose the $A(E)$ coefficient as 
$A(E) = -B(E)^*$. For convenience, we can further rewrite $B(E)$ as,
\begin{equation}
B(E) = B_0({\cal B},E) + i k B_1({\cal B},E) \>,\label{decom}
\end{equation} 
with $B_0({\cal B},E)$ and $B_1({\cal B},E)$ real for $E >0$ and both finite at $E = 0$. These coefficients can be read off equation (\ref{BEpos}). Since the solution is formally the same for $E < 0$, one can therefore analytically continue it for those values and find the bound states energies. However, since for $E < 0$ the wavenumber $k = i \sqrt{2m|E|/\hbar^2}$ is purely imaginary, the physics of the problem demands that the solution (\ref{asymp}) be finite as $r \to \infty$ and, therefore, $B(E < 0)$ must vanish, as already indicated. This is the bound state quantization condition. Denoting $E_b < 0$ as the generic values of the energies of the bound states, the condition  $B(E_b) = 0$ yields,
\begin{equation}
B_0({\cal B},E_b) - \sqrt{\frac{2m |E_b|}{\hbar^2}} B_1({\cal B},E_b) = 0 \>.\label{coupled-bound}
\end{equation}
Clearly, for negative energies $E < 0$ the coefficient $A(E < 0)$ is no longer equal to $-B^*(E< 0)$, otherwise there would not exist bound state solutions. 
It should be insisted that the  $E_b$ solutions to (\ref{coupled-bound}) are the bound state energies of the full {\it coupled} channel system, that is, they are not associated to either any of the open or closed potentials alone. \\

\begin{figure}[htbp]
\begin{center}
\includegraphics[scale=0.28]{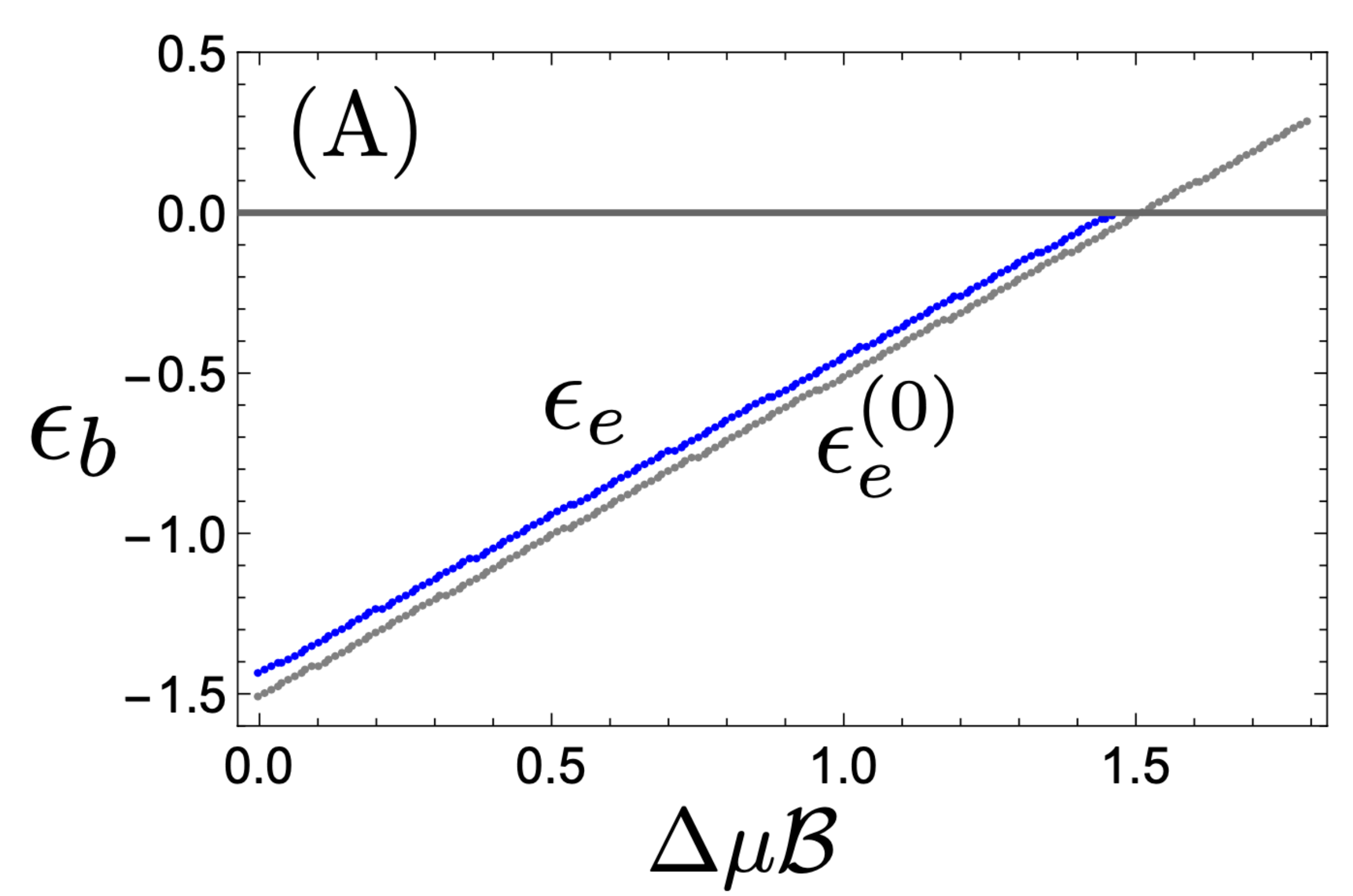}
\includegraphics[scale=0.28]{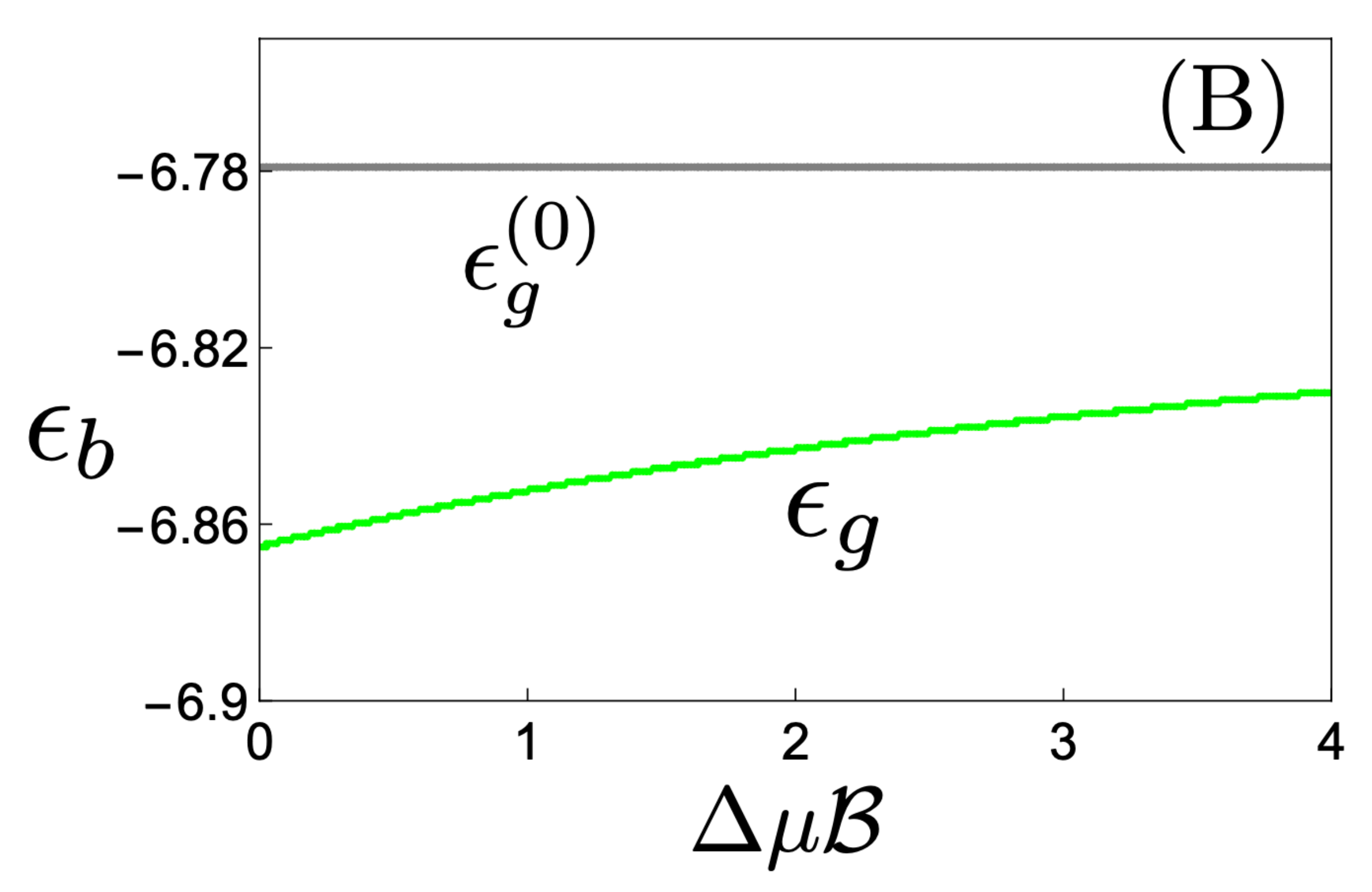}
\end{center}
\caption{(Color online) Eigenenergies of bound states of coupled and uncoupled channels as functions of $\Delta \mu {\cal B}$. (A) Coupled excited bound state  energy $\epsilon_e$ (blue) and bound state energy $\epsilon_e^{(0)}$ of the shifted uncoupled closed potential $V_c(r) + \Delta \mu {\cal B}^\star$ (gray). Notice that the coupled bound state $\epsilon_e$ no longer exists for ${\cal B} \ge {\cal B}^\star$, which is where the resonance appears. (B) Coupled ground bound state energy $\epsilon_g$ (blue) and bound state energy  $\epsilon_g^{(0)}$ of the uncoupled open potential $V_o(r)$ For this case, $U_o = 10.0$, $U_c = 4.0$, $U_{hf} = 0.75$ and $\Delta \mu {\cal B}^\star \approx 1.469$. Units $\hbar=m=R=1$.
} \label{bounds}
\end{figure}

\begin{figure}[htbp]
\begin{center}
\includegraphics[scale=0.3]{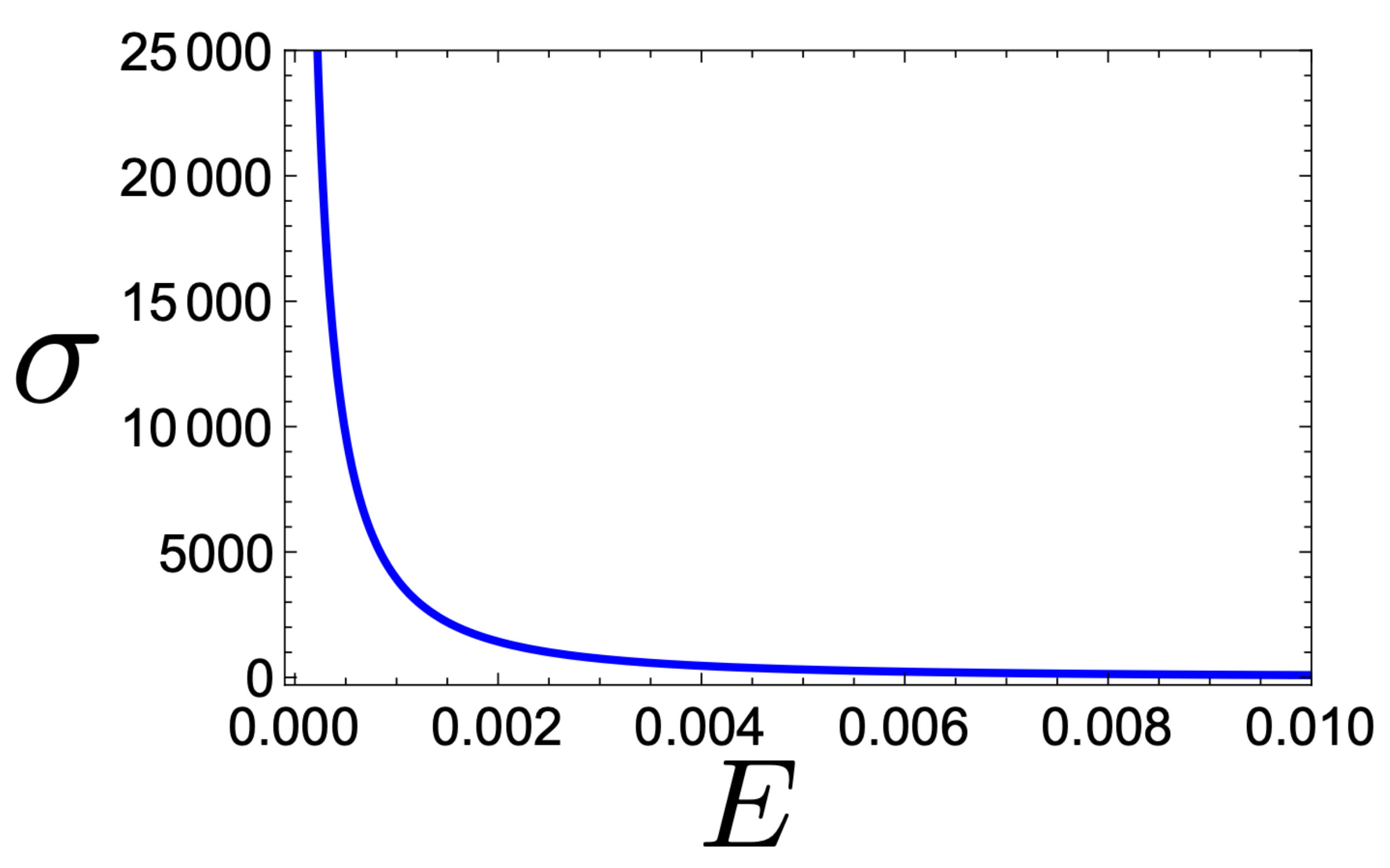}
\end{center}
\caption{(Color online) Cross section $\sigma_0$ vs incident energy $E$, at the threshold magnetic field ${\cal B}^\star$. At $E = 0$ the cross section diverges. This is the potential resonance. Units $\hbar = m = R = 1$.
} \label{sigma-res-pot}
\end{figure}

In figure \ref{bounds} we show the bound state energies of  the  particular case $U_o = 10.0 \>  \hbar^2/mR^2$, $U_c = 4.0 \> \hbar^2/mR^2$, $U_{hf} = 0.75 \> \hbar^2/mR^2$, as functions of the external field $\Delta \mu {\cal B}$. The value of $U_{hf}$ can be considered as already in the weak coupling limit. We first observe that, for this case, there are only two bound states of the coupled system, denoted as $\epsilon_g$ and $\epsilon_e$. In the figure we also plot the bound states of the respective decoupled potentials $-U_c + \Delta \mu {\cal B}$ and $U_o$, denoted as $\epsilon_g^{(0)}$ and $\epsilon_e^{(0)}$. Indeed, due to the small value of the hyperfine coupling $U_{hf}$, the true bound states are very near the decoupled bound states, however, and very crucial for our discussion, we note that the true excited bound state $\epsilon_e$ no longer exists for a threshold value ${\cal B}^\star$ of the magnetic field, indicated in the figure, while the decoupled bound state $\epsilon_e^{(0)}$ of the shifted closed potential  $-U_c + \Delta \mu {\cal B}$ certainly continues to exist. It is also relevant to point out that exactly at threshold the the bound state does not exist, since not 
not only $B(E = 0) = 0$ but also $A(E=0) = 0$. The barely existence of this bound state  at the threshold magnetic field ${\cal B}^\star$, causes a scattering resonance at incident energy $E = 0$, as shown in figure \ref{sigma-res-pot} where we plot the cross section $\sigma_0$ as a function of energy, for ${\cal B}^\star$. One finds that the cross section $\sigma_0$ grows with no bound as the incident energy vanishes $E\to 0^+$. We call this a  ``potential resonance'' since it is the exact analog of the same type of resonance that appears in single-channel scattering situations, when the single interatomic potential is shifted to the point that a bound state emerges; Landau and Lifshitz {\it Quantum Mechanics} textbook\cite{LLQM} has a very thorough discussion of this single-channel resonance that occurs at very low energies. In the ultracold gases literature it is this potential resonance of two-channel scattering at ${\cal B}^\star$ the one called a ``Feshbach resonance''.  In the following sections we will see that, contrary to the potential resonances at very low {\it coupled} bound state energies $\epsilon_b$, the resonances predicted by Feshbach occur above threshold, ${\cal B} > {\cal B}^\star$, at values of $E$ near the decoupled bound state $\epsilon_e^{(0)}$. \\

Instead of referring to the almost featureless cross section in figure \ref{sigma-res-pot}, the potential resonance is best seen in the behavior of the scattering length. For this, we use expressions (\ref{a0}) of the scattering length $a$, with the aid of the coefficients in (\ref{decom}), yielding the following explicit result,
\begin{eqnarray}
a & = & - \lim_{k\to 0} \frac{1}{k} \tan \delta_0 \nonumber \\
& = & \frac{B_1({\cal B},0)}{B_0({\cal B},0)} \>. \label{a2c} 
\end{eqnarray}

\begin{figure}[htbp]
\begin{center}
\includegraphics[width=0.5\linewidth]{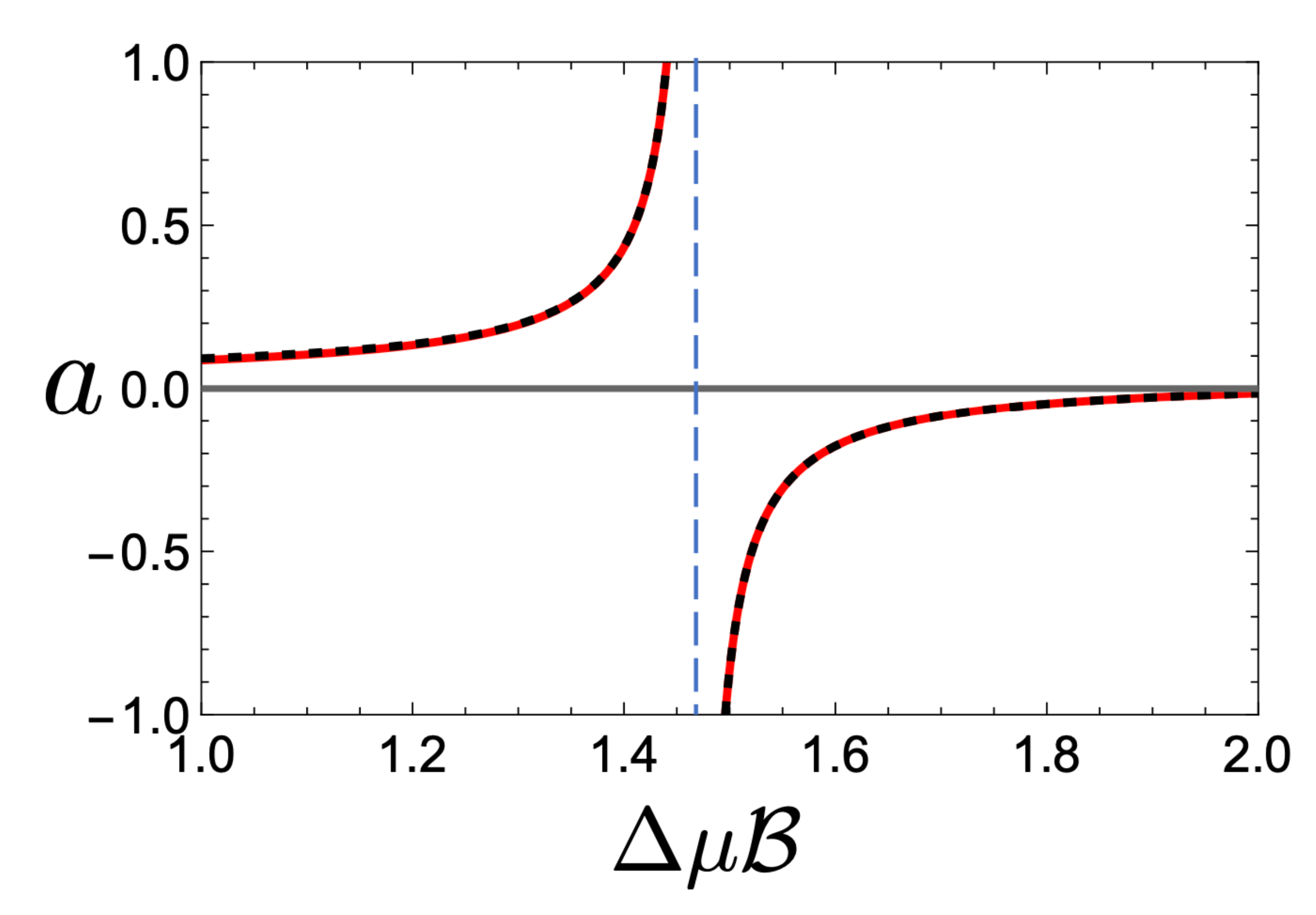}
\end{center}
\caption{(Color online) Scattering length $a$ of the full coupled-channel two-body collision as a function of $\Delta\mu {\cal B}$, for the same case as in figures \ref{bounds} and \ref{sigma-res-pot}. The (red) solid line is the exact calculation of the scattering length, evaluating  (\ref{a2c}), and the (black) dotted line is the corresponding approximated expression (\ref{coupledFeshRes2}). The vertical (black) dot-dashed line is the location of the potential resonance at $\Delta\mu {\cal B}^\star \approx 1.469$. Units $\hbar = m = R =1$
} \label{acoupled}
\end{figure}

One can analytically verify that $B_0({\cal B}^\star,0) = 0$ is the condition for having an energy eigenvalue $E_b = 0$, see (\ref{coupled-bound}), since in such a condition $B_1({\cal B}^\star,0) \ne 0$. As explained above, this can be adjusted by varying the external magnetic field to a value ${\cal B}^\star$.
For the same set of parameters as above, figure \ref{acoupled} shows the scattering length $a$ as a function of $\Delta\mu {\cal B}$, where we can observe the resonance $a \to \pm \infty$, when ${\cal B}={\cal B}^\star$. \\

It is of interest to show that the scattering length $a$, by expanding in a Taylor series $1/a$ in the vicinity of ${\cal B}^\star$, can be fit in the commonly used way in the ultracold gases literature\cite{Timmermans,Duine,Chin}. That is, 
\begin{equation}
a \approx a_{bg} \left(1 - \frac{\Delta {\cal B}}{{\cal B}-{\cal B}^\star}\right) \>,\label{coupledFeshRes2}
\end{equation}
where the  ``background'' scattering length $a_{bg}$ and the width of the resonance $\Delta {\cal B}$ are given by 
\begin{eqnarray}
a_{bg} &=& \frac{2 B_1^\prime({\cal B}^\star,0) B_0^\prime({\cal B}^\star,0)  - B_1({\cal B}^\star,0) B_0^{\prime\prime}({\cal B}^\star,0) }{2B_0^{\prime \>2}({\cal B}^\star,0) } \\
\nonumber \\
\Delta {\cal B} & = & \frac{2 B_1({\cal B}^\star,0) B_0^\prime({\cal B}^\star,0)}{ B_1({\cal B}^\star,0) B_0^{\prime\prime}({\cal B}^\star,0)- 2 B_1^\prime({\cal B}^\star,0)B_0^\prime({\cal B}^\star,0)} \>,\label{a0DelU}
\end{eqnarray}
and the primes stand for differentiation with respect to ${\cal B}$. These can be numerically evaluated and in figure \ref{acoupled} we show the comparison of this form of the scattering length with the exact expression, finding quite a good agreement. \\

Before  discussing the resonances of Feshbach theory, we make a brief but relevant detour to discuss the connection between the poles of the $S$-matrix and the resonances, in general.

\section{The poles of the $S$-matrix}

\begin{figure}[htbp]
\begin{center}
\includegraphics[scale=0.30]{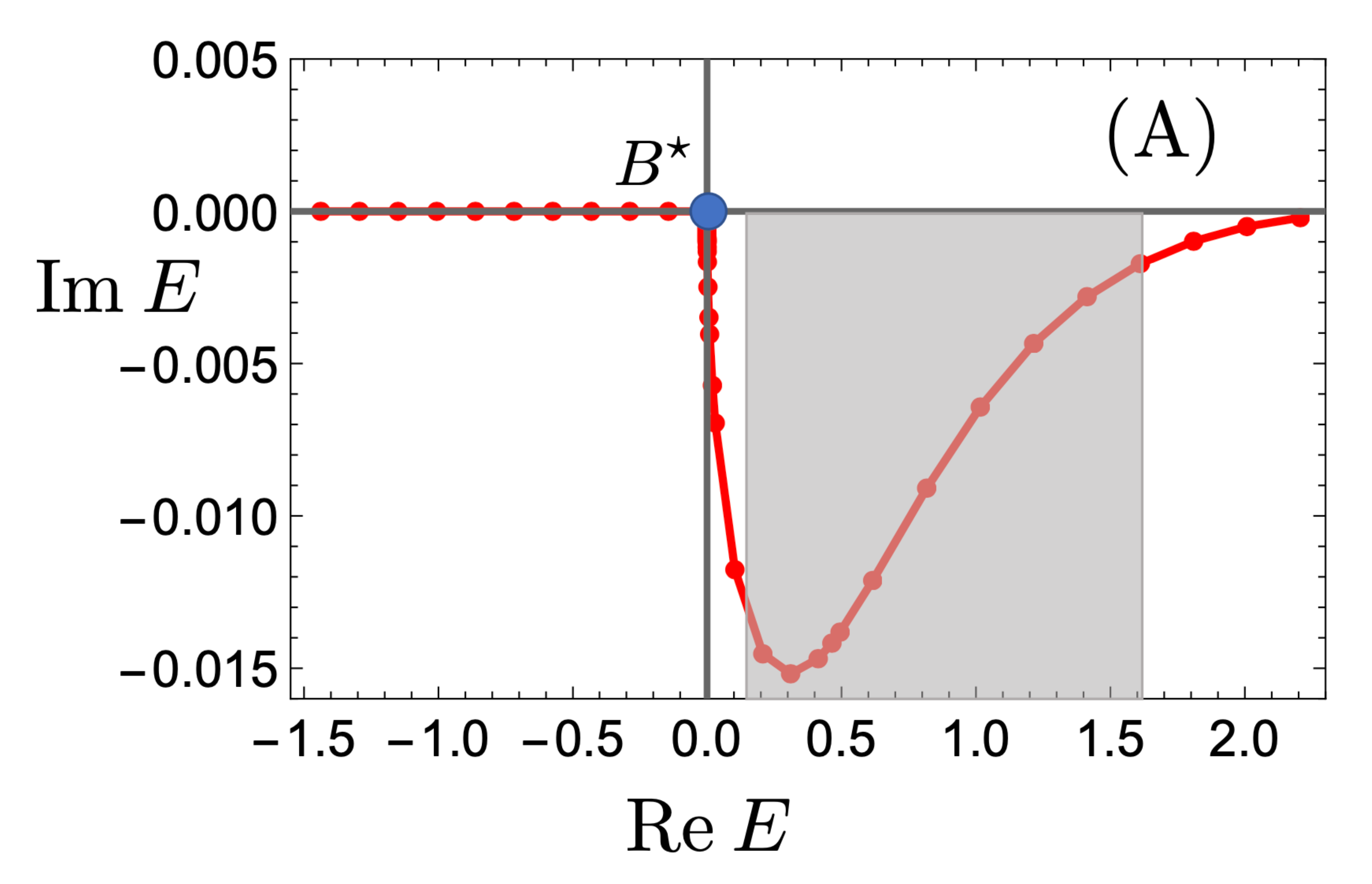}
\includegraphics[scale=0.23]{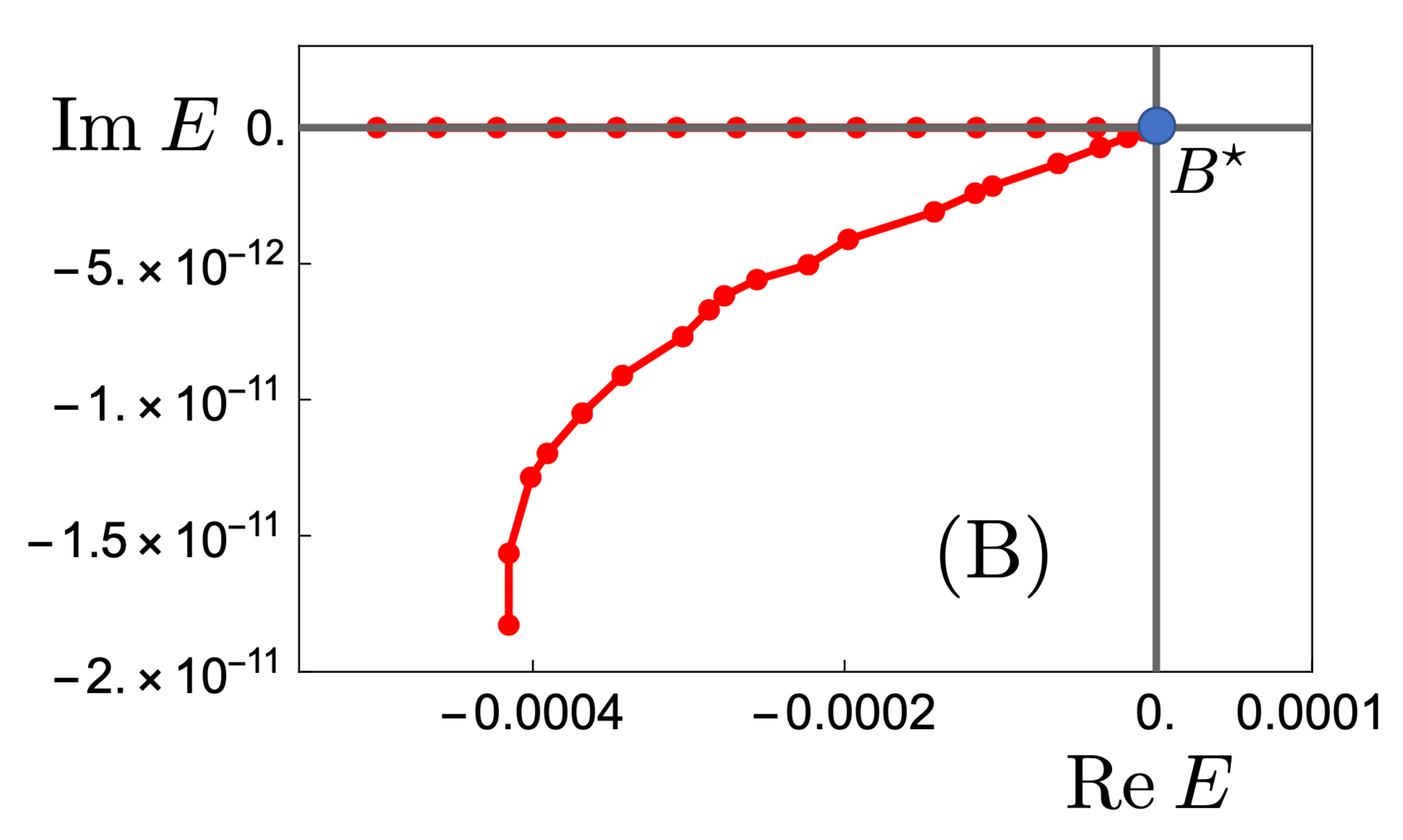}
\includegraphics[scale=0.23]{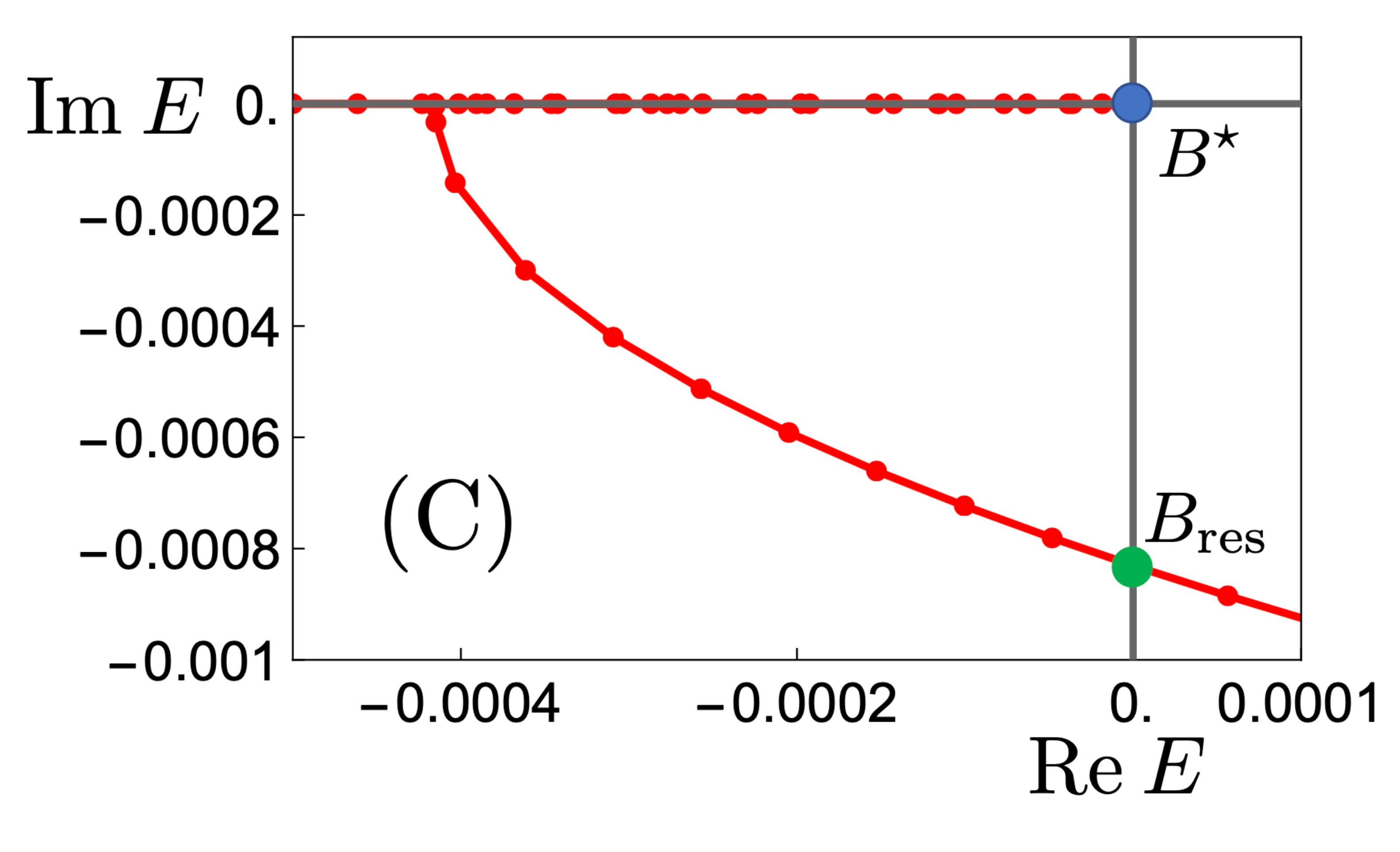}
\end{center}
\caption{(Color online) $S$-matrix poles $E_p$ in the energy $E$ complex plane for (A) all values of the external field  $0 < \Delta \mu {\cal B} < U_c$. The pole corresponding to the magnetic field ${\cal B}^\star$, yielding the potential resonance, is shown with a large (blue) dot at $E = 0$. In panel (B) we plot the very tiny region where the poles become complex with ${\rm Re}\> E_p < 0$ and ${\rm Im}\> E_p < 0$ as well. Panel (C) shows the full region of poles with ${\rm Re}\> E_p < 0$ and indicates the value ${\cal B}_{\rm res}$ of the magnetic field where the real part of the poles become positive ${\rm Re}\> E_p > 0$. Those poles produce the existence of resonance of Feshbach type. The shaded area in panel (A) is the approximated region where the resonances of Feshbach are best fitted, see Section 4. Units $\hbar = m = R = 1$.
} \label{polos}
\end{figure}

Given the simplicity of the model of square-well potentials we can calculate the simple pole structure of the $S$-matrix. As seen from expression (\ref{phashift}) $e^{2i\delta_0} = - A/B$, and already mentioned, by extending the $S$-matrix for values of the energy in the full complex plane, $E \in \mathbb{C}$, the poles of $S$ are found when $B(E) = 0$, see (\ref{BEpos}). The poles of the $S$-matrix are shown in figure \ref{polos} for the same values of the potentials considered in the previous section. The poles are found for the all the relevant values of the external magnetic field ${\cal B}$ for the closed-open channel problem at hand, namely, for the interval $0 < \Delta \mu {\cal B} < U_c$. These are fully shown in panel (A) of figure \ref{polos}; panels (B) and (C) show the very tiny region where the poles start to become complex.

As we have discussed, the bound states occur for poles $E_p$ with ${\rm Re} \> E_p < 0$ and ${\rm Im} E_p = 0$, for a given value of the magnetic field ${\cal B}$, namely $E_p = E_p({\cal B})$, see figure \ref{polos} (B). The final bound state, denoted with a large (blue) dot in the figures occurs at the threshold value ${\cal B}^*$ when the potential resonance emerges, although we insist once more that such a state does not really exist, just asymptotically. As ${\cal B}$ is increased from ${\cal B}^*$, the poles are now in the lower half $E$-plane,  ${\rm Im} \> E_p < 0$, but still with ${\rm Re} \> E_p < 0$. These are called virtual bound states since, on the one hand, the real part of its energy is below the scattering states, however, as the imaginary part is different from zero but negative, this indicates that if the system were to occupy them, those states would have a finite (and very small!) lifetime. Then, another threshold is arrived when the field equals ${\cal B} = {\cal B}_{\rm res}$, shown in a (green) large dot in panel (C) of figure \ref{polos}. Above this threshold the real part of the poles is positive, ${\rm Re} \> E_p > 0$ and ${\rm Im} E_p < 0$. These poles generate and can explain the scattering resonances, with finite values of the cross-section, that occur for $E > 0$. One  aspect to notice is that those poles are very near the axis ${\rm Im} \> E = 0$ and ${\rm Re} \> E > 0$, which is where the physical scattering takes place, or simply $E > 0$: note the very small scale in the imaginary axis in the figures. It is then found that a resonance occurs, for a fixed value of the field ${\cal B} > {\cal B}_{\rm res}$, associated to the corresponding pole. That is, based on simple geometric considerations, see Ref. \cite{Taylor}, it can be shown that as the scattering energy $E$ is moved along the real axis passing the position of the real part of the pole $E_p$, the phase shift $\delta_0$ must cross either $\pi/2$ or $3 \pi/2$, yielding a maximum of $\sigma$, see (\ref{sigma0}). As we will explicitly find, the real part of the pole is near the bound state energy of the uncoupled shifted open channel, ${\rm Re} \> E_p \approx \epsilon_e^{(0)}$. At exactly ${\cal B}_{\rm res}$ the cross section has a finite maximum at $E = 0$, with zero slope, while for  ${\cal B}^* < {\cal B} < {\cal B}_{\rm res}$ the cross section takes its largest value at $E = 0$ but with non-zero slope, suggesting that it is not a ``true'' resonance. Figure \ref{sigmas} show several examples of cross sections for different values of the external field; we will return to the discussion of such a figure in the following section after we discuss the theory of Feshbach to describe the resonances above ${\cal B}_{\rm res}$. \\

While it is clear that the resonances above the $ {\cal B}_{\rm res}$ threshold do not merge with the potential resonance at ${\cal B}^\star$, due to the tiny region of virtual bound states, we will see that the approximated theoretical expression of the resonances of Feshbach show good agreement with the exact cross section for the shaded area in figure \ref{polos} (A) only. 

\section{Resonances of Feshbach in two coupled channels}

\begin{figure}[htbp]
\begin{center}
\includegraphics[scale=0.4]{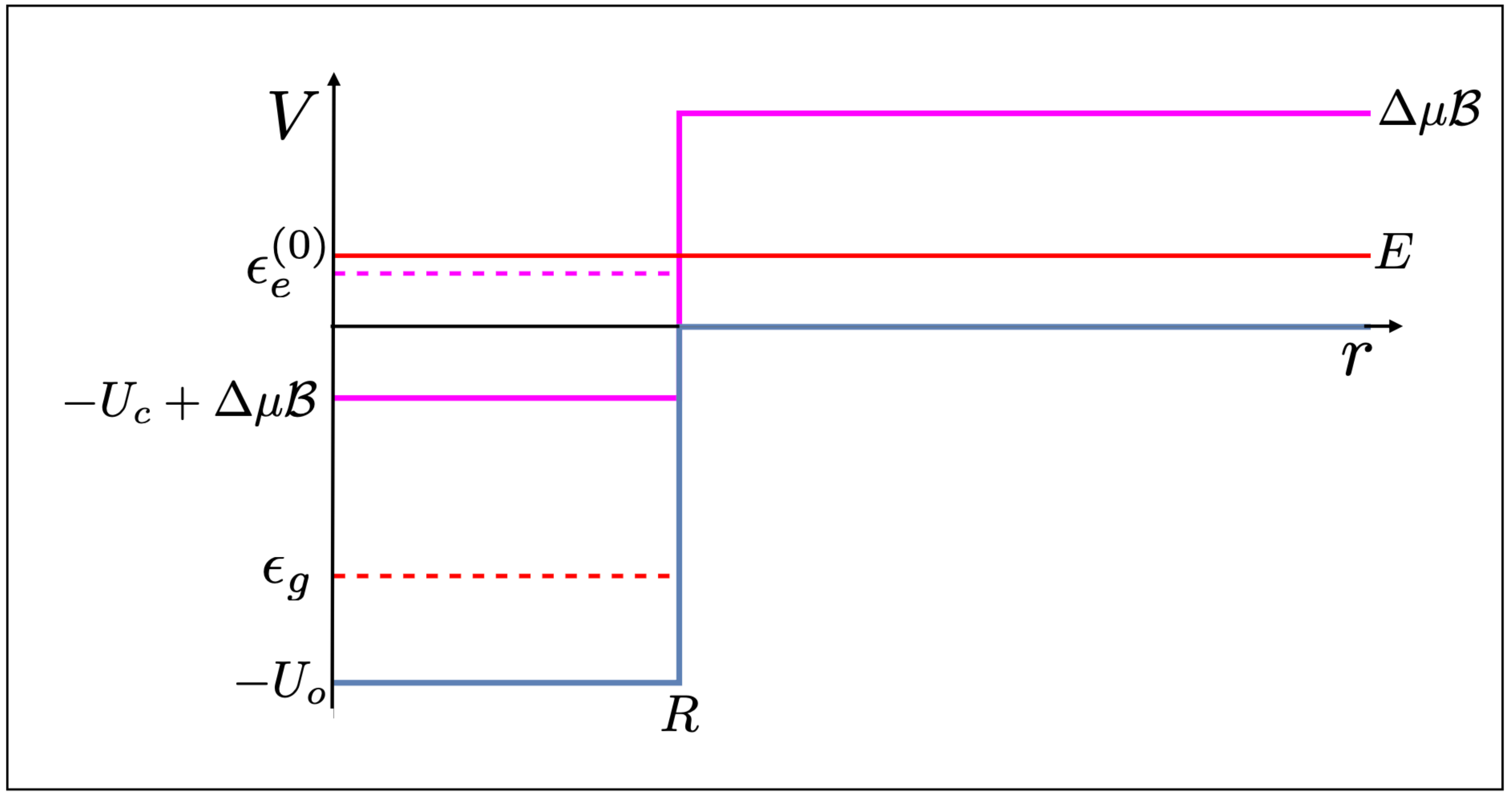}
\end{center}
\caption{(Color online) Schematic illustration of the condition for resonances of Feshbach for the  two-channel model. $-U_o$ and $-U_c + \Delta \mu {\cal B}$ are the open (blue) and shifted closed (magenta) channels. Because the magnetic field is above the threshold ${\cal B}^\star$, the energy $\epsilon_g$ corresponds to the only bound state of the full coupled system, while the state $\epsilon_e^{(0)}$ is the bound state of the {\it uncoupled} shifted closed channel. A resonance of Feshbach occurs for incident energies near such a state, namely, $E \approx \epsilon_e^{(0)}$.
} \label{esq-Fesh}
\end{figure}

In a series of papers, Herman Feshbach\cite{Fesh1,Fesh2,Fesh3} thoroughly discussed the scattering of two nucleons in multiple internal states channels, with the purpose of elucidating scattering resonances arising from the presence of those internal states in nuclear collisions. In particular, the shape of those resonances were slightly different to the well-known Breit-Wigner\cite{Breit} cross-section expression having, as shown below, a different from zero background phase-shift. While Feshbach study is quite general allowing for many different situations, the essence of the resolution of those resonances can be fairly understood with a two-channel collision, such as that described by the general Hamiltonian (\ref{Hcoupled}) in the limit of weak-interaction, namely, when the coupling term $V_{hf}$ is small compared with the depths of the open and closed channels $V_o$ and $V_c$. The resonances discussed by Feshbach can be pictured by looking at figure \ref{esq-Fesh}, exemplified with the square-well potentials model. As already argued, in such a situation, and due to the weak interaction, there appears a resonance, namely a maximum in the scattering cross section, when the incident energy $E > 0$ is very closed to the energy $\epsilon_b^{(0)} > 0$  of a bound state of the closed channel, as if it this were isolated. \\

Since the scattering occurs in the open channel, certainly influenced by the closed one, the idea of Feshbach is to reduce the two-channel scattering problem to an {\it effective} single open channel, incorporating the presence of all channels and their coupling in an approximated manner. In the following we first revisit Feshbach derivation\cite{Fesh1} of the $S$-matrix and, when applied to the simple model of square wells, we will compare the approximated Feshbach expressions for the cross section $\sigma_0$ with the exact one.\\

We first seek scattering solutions $E > 0$ to the set of Schr\"odinger equations (\ref{Hcoupled}), that we rewrite as 
\begin{eqnarray}
\hat H_o |\psi\rangle + \hat V_{hf} |\phi\rangle &=& E |\psi\rangle \nonumber \\
\hat H_c |\phi\rangle + \hat V_{hf} |\psi\rangle &=& E |\phi\rangle \>,\label{set}
\end{eqnarray}
where $\hat H_o = {\bf p}^2/2m + \hat V_o$ and $\hat H_c = {\bf p}^2/2m + \hat V_c + \Delta \hat \mu {\cal B}$, the Hamiltonians of the open and closed channels in the absence of interaction $\hat V_{hf}$. Since the scattering process in which we are interested is for $0 < E < \Delta \mu B$, given its state in the asymptotic region by $|\psi\rangle$, we formally solve for the ``closed'' state $|\phi\rangle$ from the second of the equations above,
\begin{equation}
|\phi\rangle = \frac{1}{E - \hat H_c} \hat V_{hf} |\psi\rangle \>,
\end{equation}
and substitute into the first one, yielding an equation for the scattering state $|\psi\rangle$ only:
\begin{equation}
\hat H_o |\psi\rangle  + \hat V_{hf}  \frac{1}{E - \hat H_c} \hat V_{hf} |\psi\rangle = E  |\psi\rangle \>.\label{Hefe}
\end{equation}
The presence of the closed channel in the scattering process manifests itself in the second term as an effective potential. Note that this is not truly a Schr\"odinger equation since the energy $E$ also appears in the second term. The first step is to deal with this issue. Without loss of generality we assume that the shifted closed channel Hamiltonian has a single bound state, denoted by $|\phi_0\rangle$, with energy $0 < \epsilon_0 < \Delta \mu {\cal B}$; the continuum states are $|\phi(\epsilon)\rangle$ with energy $\Delta \mu {\cal B} \le \epsilon < \infty$. These states are a complete basis set of the one-particle Hilbert state, allowing us to write (\ref{Hefe}) as,
\begin{eqnarray}
\hat H_o |\psi\rangle  &+&  \frac{1}{E - \epsilon_0}\hat V_{hf}  |\phi_0\rangle \langle \phi_0| \hat V_{hf} |\psi\rangle + \nonumber \\
&& \int_{\Delta \mu B}^\infty \hat V_{hf}  \frac{1}{E - \epsilon}|\phi(\epsilon)\rangle \langle \phi(\epsilon) | \hat V_{hf} |\psi\rangle d \epsilon = E |\psi\rangle \>.
\end{eqnarray}
Note that this is a formal solution of the problem but, we insist, the energy $E$ appears not only as an eigenvalue in the right-hand-side, but also in the denominators of the terms arising from the closed channel. As we now proceed, Feshbach theory allows us to find the $s$-wave phase shift from this expression in an appropriate approximated manner.  The main assumption, and a consideration to be applied self-consistently further below, is that the energy $E$ of the incident particle is very near the energy of the bound state $\epsilon_0$, namely $E \approx \epsilon_0$. This allows us to approximate the energy $E$ in the integral term by $\epsilon_0$ since all the values of $\epsilon^\prime$ are higher than $E$. This yields an approximated problem that we write as,
\begin{equation}
\left[ \hat H_o^{eff} + \frac{1}{E - \epsilon_0}\hat V_{hf}  |\phi_0\rangle \langle \phi_0| \hat V_{hf} \right]  |\psi\rangle \approx E  |\psi\rangle \>,\label{Hfin}
\end{equation}
where $H_o^{eff}$, being independent of $E$, is an effective Hamiltonian for the {\it open} channel:
\begin{equation}
\hat H_o^{eff} = \hat H_o + \int_{\Delta \mu B}^\infty  \frac{d \epsilon}{\epsilon_0 - \epsilon}\> \hat V_{hf}  |\phi(\epsilon)\rangle \langle \phi(\epsilon) | \hat V_{hf}  \>  \>.\label{Hefe0}
\end{equation}
While in the effective Hamiltonian  $H_o^{eff}$ the integral term, proportional to $V_{hf}^2$, will be explicitly shown to be negligible in the weak-coupling limit, the term additional to  $H_o^{eff}$ in Schr\"odinger equation (\ref{Hfin}), still depending on $E$, is essential to obtain the correct scattering properties. Let us see.\\

The scattering problem can be approached by making the usual assumption of the scattering state as an incident wave in the $z-$direction with wavevector ${\bf k}$ and its corresponding scattered part. Hence, we write $|\psi\rangle = |\Psi_{\bf k} \rangle$ in (\ref{Hfin}) and, then, we look for solutions of the form,
\begin{equation}
|\Psi_{\bf k} \rangle = | k \hat z \rangle + |\Psi_{sc, {\bf k}} \rangle \>,
\end{equation}
where
\begin{equation}
\langle {\bf r} | k \hat z \rangle = e^{ikz} \label{free}
\end{equation}
is the incident state and $|\Psi_{sc, {\bf k}} \rangle$ the scattered one. Using this we can recast Schr\"odinger equation (\ref{Hfin}) as a self-consistent solution for $|\Psi_{\bf k} \rangle$,
\begin{equation}
|\Psi_{\bf k} \rangle = |\psi_{\bf k}^+ \rangle + \frac{1}{E - \epsilon_0} \frac{1}{E^+ - \hat H_{eff}}\hat V_{hf}  |\phi_0\rangle \langle \phi_0| \hat V_{hf} |\Psi_{\bf k} \rangle \>,\label{sol1}
\end{equation}
where the label $+$ is to indicate that we are considering the out-going scattering solution. The first term $|\psi_{\bf k}^+ \rangle$ is the out-going solution to the scattering problem of the effective Hamiltonian $H_o^{eff}$  for incident energy $E$, namely,
\begin{equation}
\left(E^+ - \hat H_{eff}\right) |\psi_{\bf k}^+ \rangle = 0 \>.
\end{equation}
To verify that these two expressions are solutions to (\ref{Hfin}), simply substitute them back into it and verify that it is an identity.\\

One of the most clever steps found by Feshbach is the solution of equation (\ref{sol1}) for $|\Psi_{\bf k} \rangle$.
For this, operate in expression (\ref{sol1}) with $\hat V_{hf}$ and then project into the bound state $\langle \phi_0|$ of the closed channel; this yields an equation for $\langle \phi_0| \hat V_{hf} |\Psi_{\bf k} \rangle$ that can be solved simply:
\begin{equation}
\langle \phi_0| \hat V_{hf} |\Psi_{\bf k} \rangle = \frac{\langle \phi_0| \hat V_{hf} |\psi_{\bf k}^+ \rangle}{1- \frac{1}{E-\epsilon_0}\langle \phi_0 | \hat V_{hf} \frac{1}{E^+ - \hat H_{eff}}\hat V_{hf}  |\phi_0\rangle} \>.\label{trick}
\end{equation}
 Then, substitute this expression back into (\ref{sol1}). The final result is the full scattering state in the open channel, in terms of quantities that can in principle be found:
\begin{equation}
 |\Psi_{\bf k} \rangle = |\psi_{\bf k}^+ \rangle + \frac{1}{E^+ - \hat H_{eff}}\hat V_{hf}  |\phi_0\rangle
  \frac{\langle \phi_0| \hat V_{hf} |\psi_{\bf k}^+ \rangle}{E - \epsilon_0 - \langle \phi_0 | \hat V_{hf} \frac{1}{E^+ - \hat H_{eff}}\hat V_{hf}  |\phi_0\rangle} \>.\label{Psifin}
\end{equation}
 One should not, however, forget that this solution is for values of the energy $E$ near $\epsilon_0$, only.\\
 
 The next step is to extract the scattering $\hat S$-matrix and then limit ourselves to $s$-wave scattering. First one calculates the $\hat T$-matrix. For this we recall that if the Hamiltonian of a system is
 \begin{equation}
\hat H = -\frac{\hbar^2}{2m}\nabla^2 + \hat V
\end{equation}
then, the $\hat T$ matrix elements between two free-particle states, such as  (\ref{free}), is \cite{Newton}
\begin{equation}
\langle {\bf q} | \hat T |{\bf k} \rangle = \langle {\bf q} | \hat V |\psi_{\bf k} \rangle \>,
\end{equation}
with $|\psi_{\bf k} \rangle$ the scattering state.
Therefore, by identifying the potential $\hat V$ in Hamiltonian (\ref{Hfin}), we find the full $\hat T$-matrix of the problem, 
\begin{equation}
\langle {\bf q} | \hat T |{\bf k} \rangle = \langle {\bf q} | \left[ \hat V_{eff} + \frac{1}{E - \epsilon_0}\hat V_{hf}  |\phi_0\rangle \langle \phi_0| \hat V_{hf}\right] |\Psi_{\bf k} \rangle \>,
\end{equation}
 where $\hat V_{eff}$ can be read off (\ref{Hefe0}). Using the full solution $|\Psi_{\bf k} \rangle$, see (\ref{Psifin}), as well as the matrix element (\ref{trick}), one finds the sought for expression,
\begin{equation}
\langle {\bf q} | \hat T |{\bf k} \rangle = \langle {\bf q} | \hat T_{eff} |{\bf k} \rangle +
\frac{\langle \psi_{\bf q}^- |\hat V_{hf} | \phi_0 \rangle \langle \phi_0 |\hat V_{hf} |\psi_{\bf k}^+\rangle}{E - \epsilon_0 - \langle \phi_0 | \hat V_{hf} \frac{1}{E^+ - \hat H_{eff}}\hat V_{hf}  |\phi_0\rangle} \>,
\end{equation}
where $\hat T_{eff}$ is the $\hat T$-matrix of the effective {\it open} Hamiltonian (\ref{Hefe0}); and the state $\langle \psi_{\bf q}^- |$ is the {\it in-going} scattering state of the same effective Hamiltonian, given by,
\begin{equation}
\langle \psi_{\bf q}^- | = \langle {\bf q}| + \langle {\bf q}| \hat V_{eff} \frac{1}{E^+ - \hat H_{eff}} \>.
\end{equation}

The $\hat S$-matrix is related to the $\hat T$-matrix as,
\begin{equation}
\hat T = \frac{1}{2\pi i}\left(\hat 1 - \hat S\right) \>, 
\end{equation}
and, therefore, one obtains
\begin{equation}
\langle {\bf q} | \hat S |{\bf k} \rangle = \langle {\bf q} | \hat S_{eff} |{\bf k} \rangle 
- 2 \pi i\frac{\langle \psi_{\bf q}^- |\hat V_{hf} | \phi_0 \rangle \langle \phi_0 |\hat V_{hf} |\psi_{\bf k}^+\rangle}{E - \epsilon_0 - \langle \phi_0 | \hat V_{hf} \frac{1}{E^+ - \hat H_{eff}}\hat V_{hf}  |\phi_0\rangle} \>,\label{Sbig}
\end{equation}
In order to relate the $\hat S$-matrix to the scattering phase shifts, one considers the on-shell solution ${\bf q} \to {\bf k}$. In the $s$-wave approximation one has \cite{Newton}
\begin{equation}
\langle {\bf q} | \hat S |{\bf k} \rangle \approx e^{2i\delta_0} \>\>\>\>{\bf q} \to {\bf k}
\end{equation}
with $\delta_0$ the $s$-wave phase shift of the full problem, and analogously for $\hat S_{eff}$, with a corresponding $s$-wave phase shift of the effective open channel. Furthermore, recalling the relation between in-going and out-going  scattering solutions \cite{Newton}
\begin{equation}
\langle \psi_{\bf k}^- | \approx e^{2i\delta_0^{eff}} \langle \psi_{\bf k}^+ | \>,
\end{equation}
one finds, in (\ref{Sbig}),
 \begin{equation}
e^{2i\delta_0} \approx e^{2i\delta_0^{eff}}\left[1
- 2 \pi i\frac{\langle \psi_{\bf k}^+ |\hat V_{hf} | \phi_0 \rangle \langle \phi_0 |\hat V_{hf} |\psi_{\bf k}^+\rangle}{E - \epsilon_0 - \langle \phi_0 | \hat V_{hf} \frac{1}{E^+ - \hat H_{eff}}\hat V_{hf}  |\phi_0\rangle} \right]\>.
\end{equation}
The last step is to recall that the matrix elements above are to be evaluated at $E=\epsilon_0$ and $E^+ = \epsilon_0 + i \varepsilon$ and then taking the limit $\varepsilon \to 0^+$ \cite{delta}. By renaming $\delta_{bg} = \delta_0^{eff}(\epsilon_0)$ as the background phase-shift, we arrive to the final expression as found by Feshbach,
\begin{equation}
e^{2i\delta_0} \approx e^{2i\delta_{bg}}\left[1
- \frac{i \pi \Gamma}{E - ( \epsilon_0 + \Delta \epsilon_0) + i \pi \frac{\Gamma}{2}} \right]\>.\label{Fesh}
\end{equation}
 This indicates that the resonance is at very near the position of the pole $E_p = ( \epsilon_0 + \Delta \epsilon_0) - i \pi \frac{\Gamma}{2}$, where
\begin{equation}
\frac{\Gamma}{2} = \left| \langle \phi_0 |\hat V_{hf} |\psi_{{\bf k}_0}^+\rangle \right|^2 \>\>\>\>\>\> |{\bf k}_0| = \sqrt{\frac{2m \epsilon_0}{\hbar^2}} \>\label{Gamma}
\end{equation}
is the resonance width, and the energy shift of the resonance is given by
\begin{equation}
\Delta \epsilon_0 = \langle \phi_0 | \hat V_{hf} {\cal P}\frac{1}{\epsilon_0 - \hat H_{eff}}\hat V_{hf}  |\phi_0\rangle \>,\label{Deltae0}
\end{equation}
with ${\cal P}$ denoting the principal value. It is of relevance to observe that, in order to calculate these coefficients, we need to know {\it all} the states of the shifted closed channel $V_c(r) + \Delta \mu {\cal B}$ and of the effective open channel given by (\ref{Hefe0}). In particular, for the calculation of the principal value in the evaluation of $\Delta \epsilon_0$, one must use the full basis set of states of $\hat H_{eff}$ including the states in the continuum where the energy $\epsilon_0$ lies. In Appendix B we give details of how the parameters of Feshbach $S$-matrix are calculated and, in particular, how the weak-limit approximation becomes useful. To summarize, from expression (\ref{Fesh}) one finds the $s$-wave phase shift $\delta_0$ for energies $E$ of the open channel, in the vicinity of the bound state energy $\epsilon_0$ of the shifted closed channel. The scattering cross section $\sigma_0$ (\ref{sigma0}) can therefore be found. \\

\begin{figure}[htbp]
\begin{center}
\includegraphics[scale=0.2]{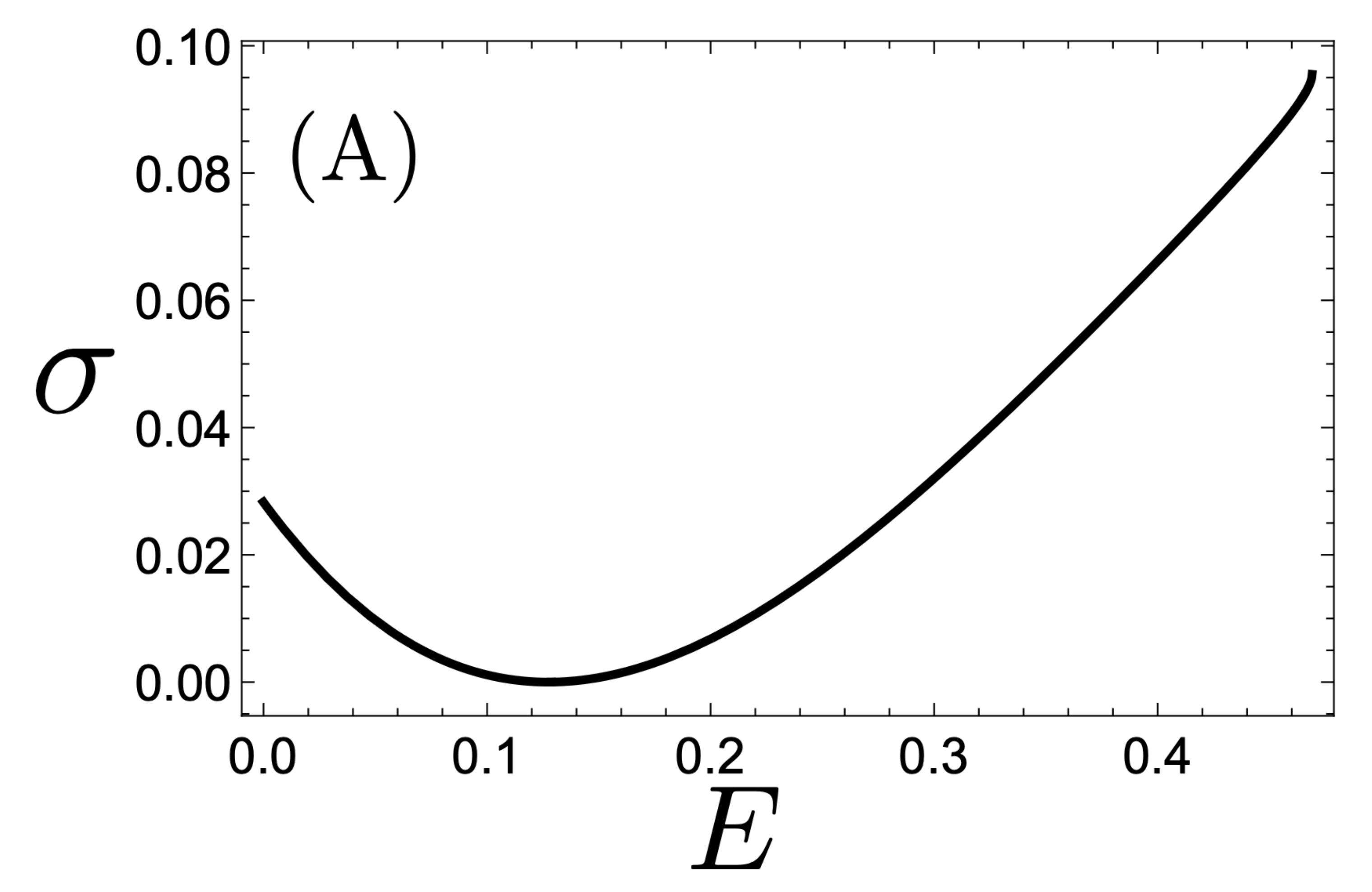}
\includegraphics[scale=0.2]{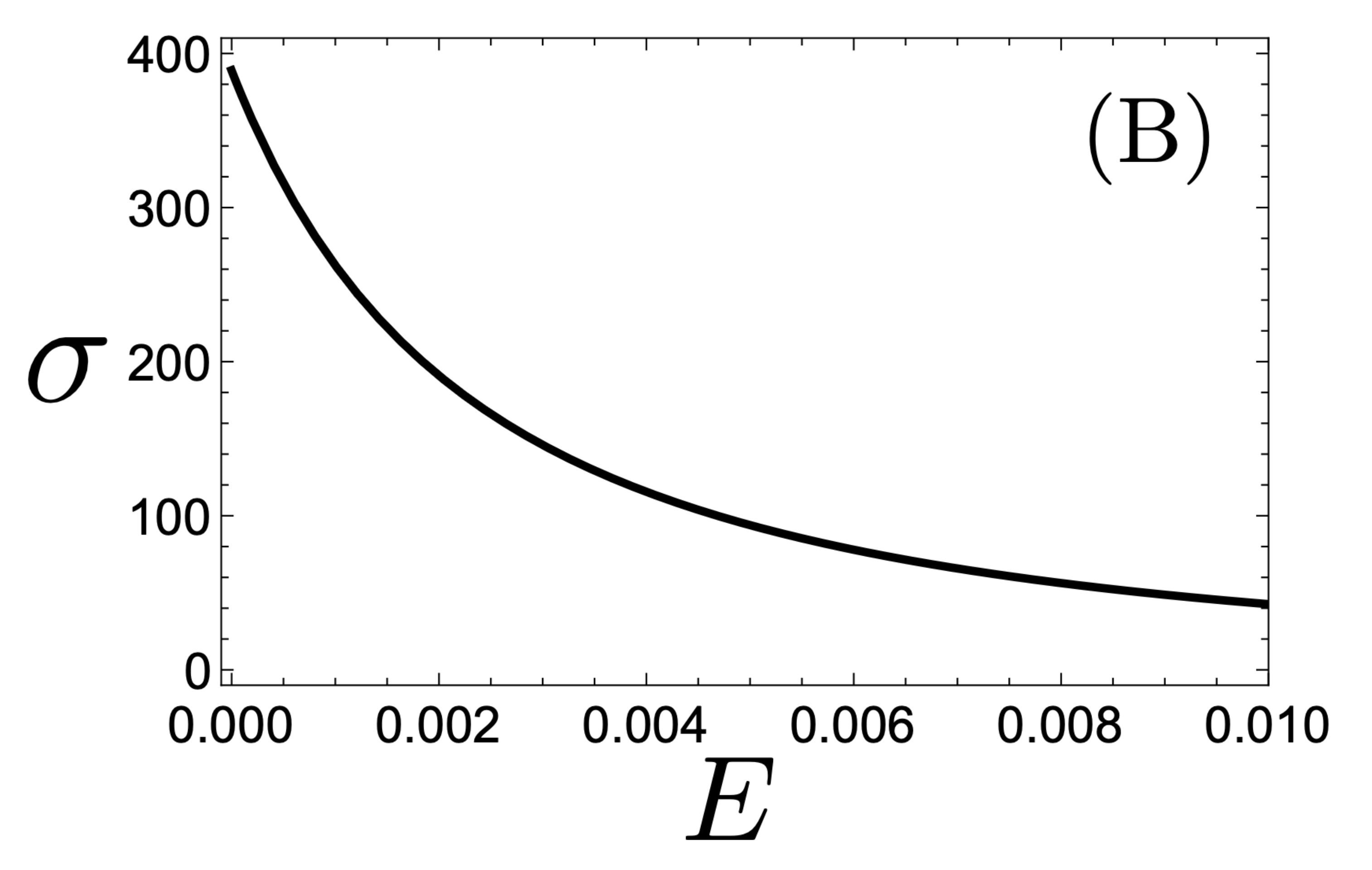}
\includegraphics[scale=0.27]{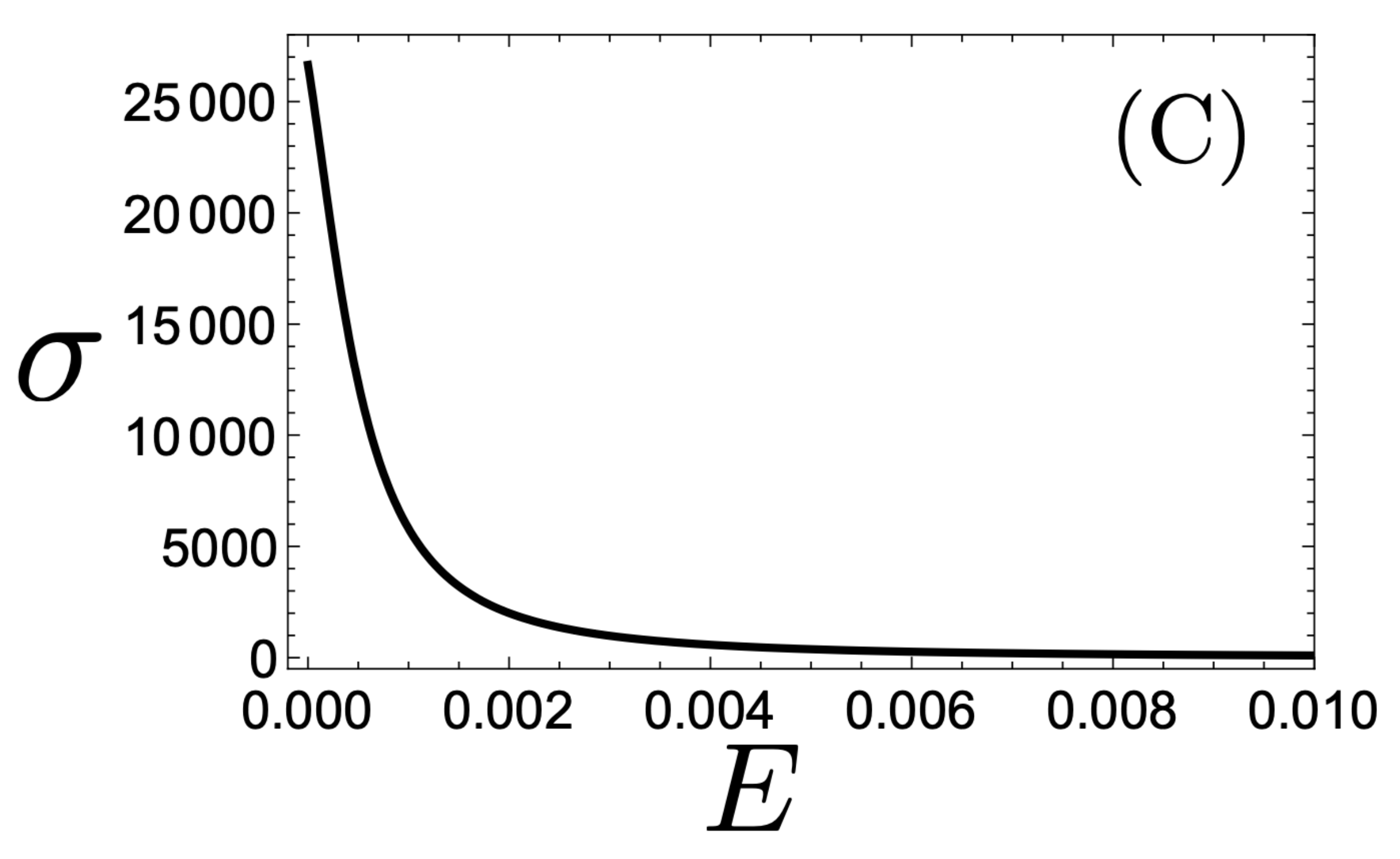}
\includegraphics[scale=0.25]{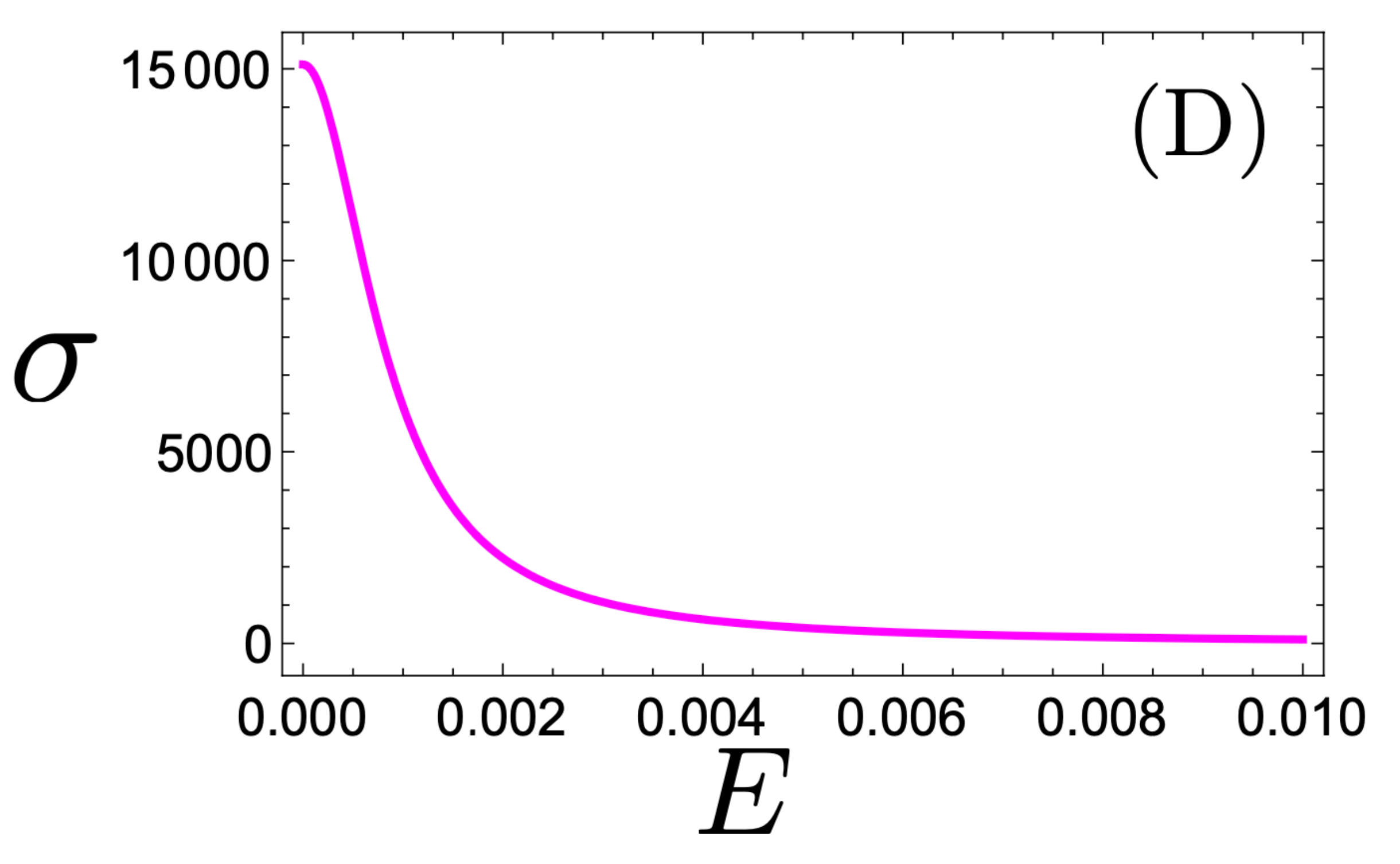}
\includegraphics[scale=0.25]{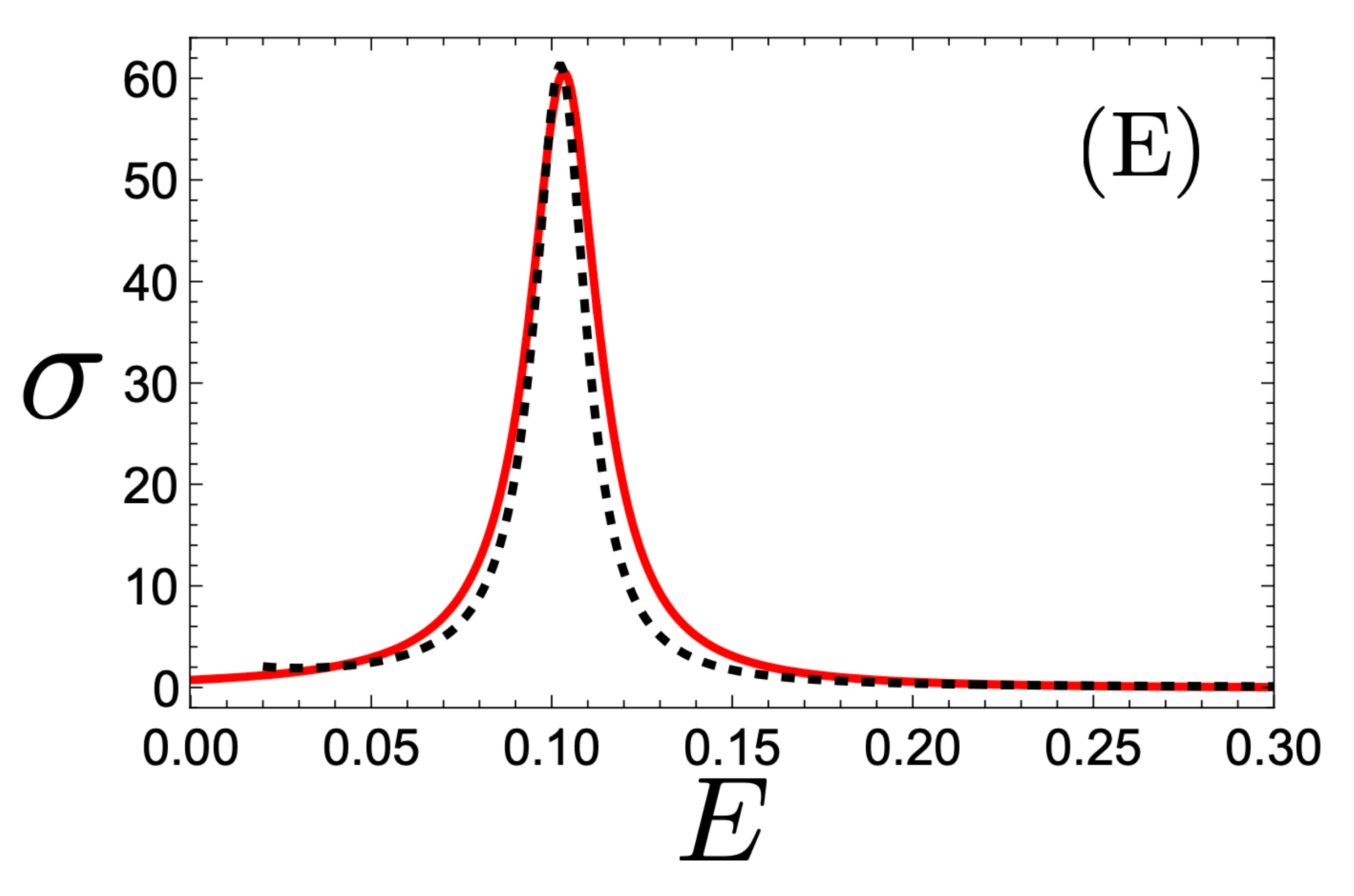}
\includegraphics[scale=0.25]{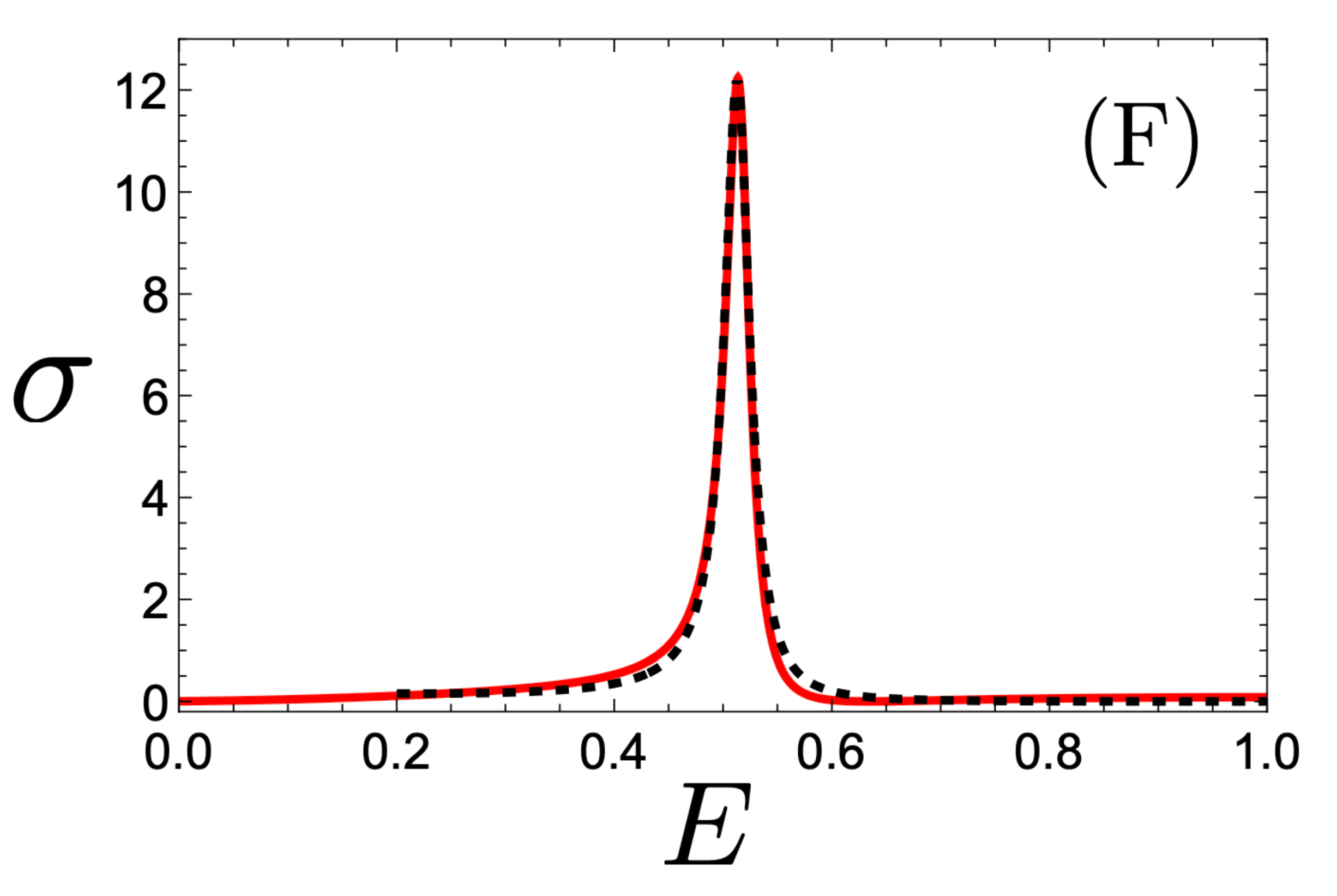}
\includegraphics[scale=0.25]{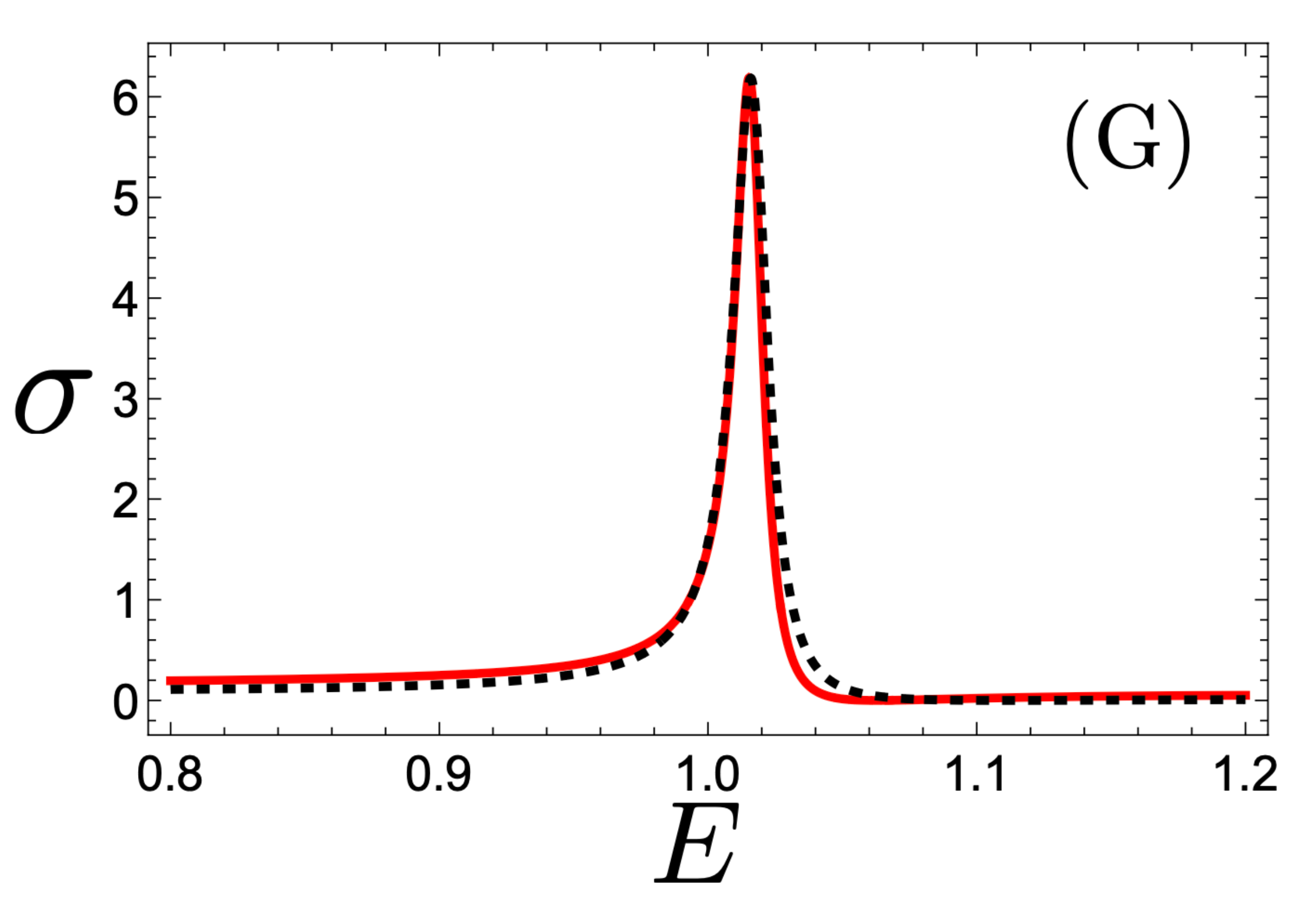}
\includegraphics[scale=0.25]{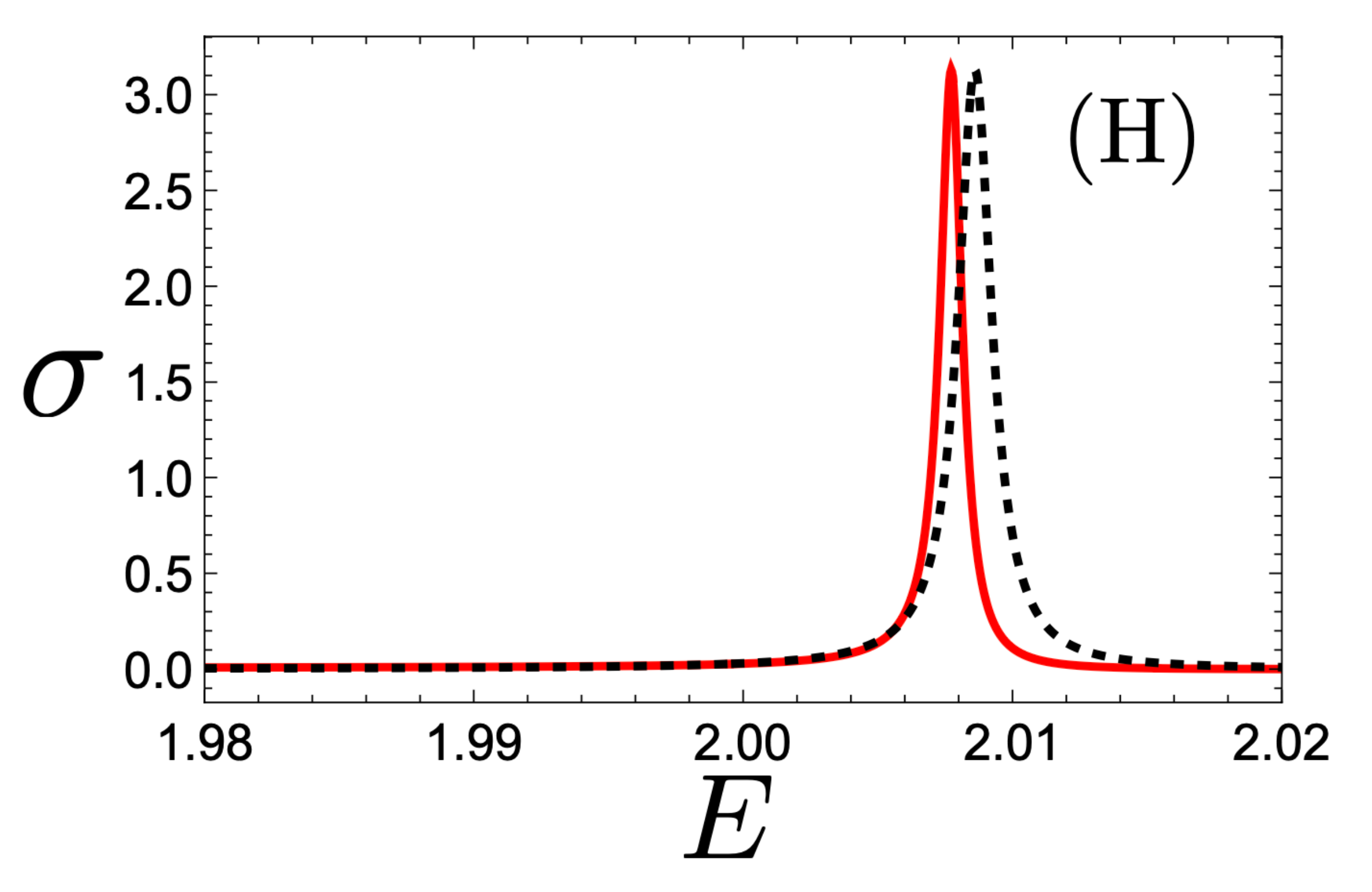}
\end{center}
\caption{(Color online) Cross section $\sigma$ vs incident energy $E$, for different values of the external field ${\cal B}$. See Section 5 for the description of each panel. In panels (E) to (H) we show comparisons between exact numerical evaluation of the cross section, solid (red) line, and Feshbach approximation in dotted (black) lines.The values of the background phase shift $\delta_{bg}$, the resonance width $\Gamma/2$ and the energy shift $\Delta \epsilon_0$ are given in Table I. Units $\hbar = m = R = 1$.
} \label{sigmas}
\end{figure}

Figure \ref{sigmas} illustrates and summarizes our discussion. It shows the cross section $\sigma_0$, see (\ref{sigma0}), with the exactly (numerically) calculated $S$-matrix (\ref{phashift}) and, where appropriate, the approximated form (\ref{Fesh}) of the theory of Feshbach, for different values of the external field ${\cal B}$, using the same values as above for the square well parameters.  Panels (A) and (B) show cross sections for values of the magnetic field below the threshold ${\cal B}^*$; the potential resonance at exactly ${\cal B}^*$ is shown in figure \ref{sigma-res-pot}; (C) for a value ${\cal B}^* < {\cal B} < {\cal B}_{\rm res}$; (D) at the threshold ${\cal B}_{\rm res}$; and (E) to (H) at resonances above the previous thresholds, including the Feshbach prediction in dashed (black) lines. We can see that the agreement  is very impressive for intermediate values, as those indicated in the shaded area of figure \ref{polos} (A). There is an additional and final observation: although not evident in all panels, all cross sections, except very near ${\cal B} = 0$, show another type of resonance, namely, the cross-section $\sigma_0$ always has an energy where it vanishes. This means that the phase-shift is $\delta = \pi$ at those energies. This is also a resonance in the sense that for such an energy the atoms become ``transparent'' to each other. While this is expected from Feshbach expression, and it is a known phenomenon, it is interesting that it always occurs. \\

We stress that the Feshbach cross-sections shown in figure \ref{sigmas} are not fittings of his formula (\ref{Fesh}) but true calculations of the parameters $\delta_{bg}$, $\Gamma$ and $\Delta \epsilon_0$, for a given $\epsilon_0$ which, in turn, depends on the given value of the magnetic field ${\cal B}$; these values are shown in Table I. While these are not calculations of extraordinary difficulty, they do require of a careful evaluation and assessment of the full problem of the effective open Hamiltonian $\hat H_o^{eff}$. The details of these calculations are given in Appendix B.\\
 
\begin{table}[h!]
\centering
\begin{tabular}{|| c |c |c |c |c |c ||} 
 \hline
 & $\> \Delta \mu {\cal B} \>$ & $\> \delta_{bg} \>$ & $\> \Gamma/2 \>$ & $\> \epsilon_0 \>$ & $\> \Delta \epsilon_0 \>$ \\ 
 \hline\hline
(E) & 1.57 & -0.023 & 0.0028 & 0.06 & 0.042\\ 
 (F) & 1.97 & 0.025 & 0.0045 & 0.46 & 0.053 \\
 (G) & 2.47 & 0.082 & 0.0025 & 0.96 & 0.056 \\
 (H) & 3.47 & 0.001 & 0.00024 & 1.96& 0.048  \\
 \hline
\end{tabular}
\caption{Calculated values of the bound state energy of the shifted closed channel with energy $\epsilon_0$; background phase shift $\delta_{bg}$, resonance width $\Gamma/2$ and energy shift $\Delta \epsilon_0$, for different values of the Zeeman external field $\Delta \mu {\cal B}$ for panels (E)-(H) of figure \ref{sigmas}. See Appendix B for details. Units $\hbar = m = R = 1$.}
\label{table:1}
\end{table}

\newpage

\section{Final Remarks}

While we believe the previous sections are self-contained and enough for the purposes of the article, two additional brief remarks can be made. One is the fact that we have limited the discussion to the weak coupling limit $U_{hf} \ll U_o, U_c$ because it is in this situation that Feshbach theory can be fairly applied. However, the exact solution of this simple model, albeit numerical, can be performed for any value of the parameters allowing for an exploration of all sorts of different physical situations. And second, partly a follow up of the last statement, we would like to insist in the additional pedagogical value that this model can have in the teaching of scattering theory. We mention here that all the numerical evaluations of the article were mostly made\cite{clarify} with the use of {\it Mathematica}\cite{Math}, the main technical difficulty of the elucidation of the different expressions being the personal ability of finding reduced analytical expressions that can be, then, numerically evaluated. In this regard, in addition to being useful in the understanding of the physics of coupled channel scattering, we believe that the present study can be implemented as a special topic in an advanced undergraduate or graduate lecture course on quantum mechanics. 

\ack 
We thank PAPIIT IN117623 UNAM grant for support. GAS acknowledges a scholarship from CONACYT-Mexico.

\appendix
\setcounter{section}{0}

\section{Solution to the coupled two-channel model with square well potentials}

Consider Schr\"odinger equation (\ref{Hcoupled}) with the square-well potentials given in (\ref{potentials}), with range $R$. To find the solution we separate the problem into two, (I) for $0 \le r < R$
\begin{equation}
\fl \left(\begin{array}{cc}
-\frac{\hbar^2}{2m} \frac{d^2}{dr^2} - U_o & U_{hf}  \\
U_{hf} & -\frac{\hbar^2}{2m} \frac{d^2}{dr^2} - U_c + \Delta \mu {\cal B}
\end{array}\right)
\left(\begin{array}{c}
u_I(r) \\
v_I(r)\end{array}\right) = E \left(\begin{array}{c}
u_I(r) \\
v_I(r)\end{array}\right) \>,\label{SchrLR}
\end{equation}
and (II) for $r > R$,
\begin{equation}
\fl \left(\begin{array}{cc}
-\frac{\hbar^2}{2m} \frac{d^2}{dr^2}  & 0 \\
0 & -\frac{\hbar^2}{2m} \frac{d^2}{dr^2}  + \Delta \mu {\cal B}
\end{array}\right)
\left(\begin{array}{c}
u_{II}(r) \\
v_{II}(r)\end{array}\right) = E \left(\begin{array}{c}
u_{II}(r) \\
v_{II}(r)\end{array}\right) \>,\label{SchrGR}
\end{equation}
for given parameters $U_o > U_c>0$, $0 < \Delta \mu {\cal B} < U_c$, and $U_{hf} \ll U_c$. We seek solutions for $0 < E < \Delta \mu{\cal B}$. The wavefunctions $u(r)$ and $v(r)$ and their derivatives are continuous at $r = R$. The solution (I) is zero at $r = 0$. One way to proceed is, first, to write the $2 \times 2$ matrix problem for $r < R$ in diagonal form, since the solution  is thus easier, and then return to the the initial basis to make the matching at $r = R$. Hence, we first find the (constant) rotation matrix $U(\theta)$ that diagonalizes the constant part of the Hamiltonian for $r < R$  (\ref{SchrLR}),
\begin{equation}
U(\theta) \left(\begin{array}{cc}
 - U_o & U_{hf}  \\
U_{hf} &  - U_c + \Delta \mu {\cal B}
\end{array}\right)U^{-1}(\theta) = \left(\begin{array}{cc}
 - \epsilon_{+} & 0 \\
0 & -\epsilon_{-}
\end{array}\right) \>.
\end{equation}
with
\begin{equation}
U(\theta) =\left(\begin{array}{cc}
\cos \frac{\theta}{2} & \sin \frac{\theta}{2}  \\
-\sin \frac{\theta}{2} &  \cos \frac{\theta}{2}
\end{array}\right) \>.
\end{equation}
The values of  $\epsilon_{\pm}$  and $\theta$ are given in eqs. (\ref{epm}) and (\ref{teta}). With the restrictions imposed above we ensure that $\epsilon_{\pm} > 0$. Applying this rotation matrix to (\ref{SchrLR}) for $r < R$ leads to the simpler equation
\begin{equation}
\left(\begin{array}{cc}
-\frac{\hbar^2}{2m} \frac{d^2}{dr^2} - \epsilon_{+} & 0  \\
0 & -\frac{\hbar^2}{2m} \frac{d^2}{dr^2} - \epsilon_{-}
\end{array}\right)
\left(\begin{array}{c}
\tilde u_I(r) \\
\tilde v_I(r)\end{array}\right) = E \left(\begin{array}{c}
\tilde u_I(r) \\
\tilde v_I(r)\end{array}\right) \>,\label{SchrLRrot}
\end{equation}
whose solution is
\begin{equation}
\left(\begin{array}{c}
\tilde u_I(r) \\
\tilde v_I(r)\end{array}\right) =
\left(\begin{array}{c}
D(E) \sin k_+ r \\
F(E) \sin k_ - r \end{array}\right) \>,
\end{equation}
with $k_{\pm} = \sqrt{2m \epsilon_{\pm}/\hbar^2}$, see eqs. (\ref{ks}). The coefficients $D(E)$ and $F(E)$, functions of $E$, are to be determined as explained below. Thus, the solution for $r < R$ is found as,
\begin{equation}
\left(\begin{array}{c}
 u_I(r) \\
 v_I(r)\end{array}\right) = U^{-1}(\theta)
\left(\begin{array}{c}
\tilde u_I(r) \\
\tilde v_I(r)\end{array}\right) \>. \label{solLR}
\end{equation}

The solution for $r > R$ of (\ref{SchrGR}) is straightforwardly given by,
\begin{equation}
\left(\begin{array}{c}
u_{II}(r) \\
v_{II}(r)\end{array}\right) = 
\left(\begin{array}{c}
A(E) e^{ikr} + B(E) e^{-ikr} \\
C(E) e^{-\kappa_B r}\end{array}\right)
\> ,\label{solGR}
\end{equation}
with $k = \sqrt{2mE/\hbar^2}$ and $\kappa_B = \sqrt{2m(\Delta \mu {\cal B}-E)/\hbar^2}$. With the matching of the wavefunctions (\ref{solLR}) and (\ref{solGR}), and their derivatives, at $r = R$, after a lengthy exercise, we can find the coefficients $A, B, C, D$ and $F$. In this way, we obtain the $S$-matrix $e^{2i\delta_0} = - A(E)/B(E)$, with $B(E)$ as in (\ref{BEpos}); the overall real constant ${\cal C}_0$ can be further found by ortho-normalizing the solutions in the continuum of energies. The bound states and their eigenenergies can also be found by analytically continuing the positive energy solution here presented for $E > 0$, to negative values $E < 0$ and imposing $B(E< 0) = 0$, since $k$ becomes purely imaginary, $k = i \sqrt{2m|E|/\hbar^2}$, see (\ref{solGR}).

\section{Numerical evaluation of Feshbach $S$-matrix}

The purpose of this Appendix is to show the details of the calculations, as well as the approximations used, of the background phase shift  $\delta_{bg}$, the resonance width $\Gamma$ and the energy shift $\Delta \epsilon_0$, that appear in the $S$-matrix of Feshbach theory, given in (\ref{Fesh}), (\ref{Gamma}) and (\ref{Deltae0}). As one can observe from those expressions, one needs {\it all} the bound and scattering states, properly normalized, of both, the shifted closed channel and the effective open channel.\\

First, as it will be shown to be useful below, we consider the full solution of a {\it single} channel with a square-well potential
\begin{equation}
-\frac{\hbar^2}{2m} \frac{d^2}{dr^2} u(r) + V(r) u(r) = E u(r) \>, \label{Sch1}
\end{equation}
\begin{equation}
V(r) =\left\{
\begin{array}{ccc}
- U &\textrm{if} & r \le R \\
 & & \\
 0 & \textrm{if} & r > R 
 \end{array} \right. \>.
 \end{equation}
 Limiting ourselves to the $s$-wave case, $l = 0$, we assume with no loss of generality that $V(r)$ holds a single bound state $|\phi_b \rangle$ with eigenenergy $\epsilon_b$. These states, including the bound state, are a complete base for $l = 0$ that can be properly normalized. The bound state, $\phi_b(r) = \langle {\bf r} | \phi_b \rangle$, is
\begin{equation}
\phi_b(r) = \left\{
\begin{array}{ccc} 
\frac{1}{\sqrt{4 \pi} r} \sqrt{\frac{2\kappa_b}{\kappa_b R +1}} \sin k_b r & {\rm for} & r \le R \\
 & & \\
\frac{1}{\sqrt{4 \pi} r} \sqrt{\frac{2\kappa_b}{\kappa_b R +1}}e^{-\kappa_b(r-R)}  \sin k_b R & {\rm for} & r \ge R 
\end{array} \right. \>,\label{boundst}
\end{equation}
where
\begin{equation}
\kappa_b = \sqrt{\frac{2 m|\epsilon_b|}{\hbar^2}} \>\>\>\>\>\>\>\>\>\> k_b = \sqrt{\frac{2 m (U - |\epsilon_b|)}{\hbar^2}} \>.
\end{equation}
The matching of the first derivative of $u_b(r) = r \phi_b(r)$ at $r = R$ yields the quantization condition for the energy $\epsilon_b$,
$k_b \cos k_b R = - \kappa_b \sin k_b R$. The states of the continuum are, $\phi(\epsilon,r) = \langle {\bf r} | \phi(\epsilon) \rangle$
\begin{equation}
\phi(\epsilon,r) =\left\{
\begin{array}{ccc} 
\frac{-2ik}{|\xi|  r} \sqrt{\frac{m}{8\pi^2 \hbar^2 k}} \sin k_I r & {\rm for} & r \le R \\
 & & \\
\frac{1}{|\xi| r}  \sqrt{\frac{m}{8\pi^2 \hbar^2 k}} \left[ - \xi e^{ik(r-R)}+\xi^* e^{-ik(r-R)}\right] & {\rm for}& r \ge R 
\end{array} \right. \>,\label{scatst}
\end{equation}
where
\begin{eqnarray}
k &=& \sqrt{\frac{2 m \epsilon}{\hbar^2}} \nonumber \\
k_I &=& \sqrt{\frac{2 m (U +\epsilon)}{\hbar^2}}
\end{eqnarray}
and
\begin{equation}
\xi = k_I \cos k_I R + i k \sin k_I R \>.
\end{equation}
The bound state $\phi_b(r)$ is normalized to one and it is orthogonal to the continuum states $\phi(\epsilon,r)$. The latter are normalized as,
\begin{equation}
\int d^3 r \> \phi^*(\epsilon,r) \phi(\epsilon^\prime,r) = \delta(\epsilon - \epsilon^\prime) \>.
\end{equation}
 
To obtain the states and energies of the shifted closed potential, we consider that the external magnetic field ${\cal B}$ is such that the energy of the bound state $\epsilon_0$ is within $0 < \epsilon_0 < \Delta \mu {\cal B}$. The states of the continuum $|\phi(\epsilon)\rangle$ are limited also  by $\Delta \mu {\cal B} < \epsilon < \infty$. The corresponding solutions are found from the above formulae by writing $|\epsilon_b| = \Delta \mu {\cal B} - \epsilon_0$, $U - |\epsilon_b|= U_c -  \Delta \mu {\cal B} + \epsilon_0$ for the bound state. For the continuum states $k$ and $k_I$ remain as such, with $U = U_c- \Delta \mu {\cal B}$. 

\begin{figure}[htbp]
\begin{center}
\includegraphics[scale=0.35]{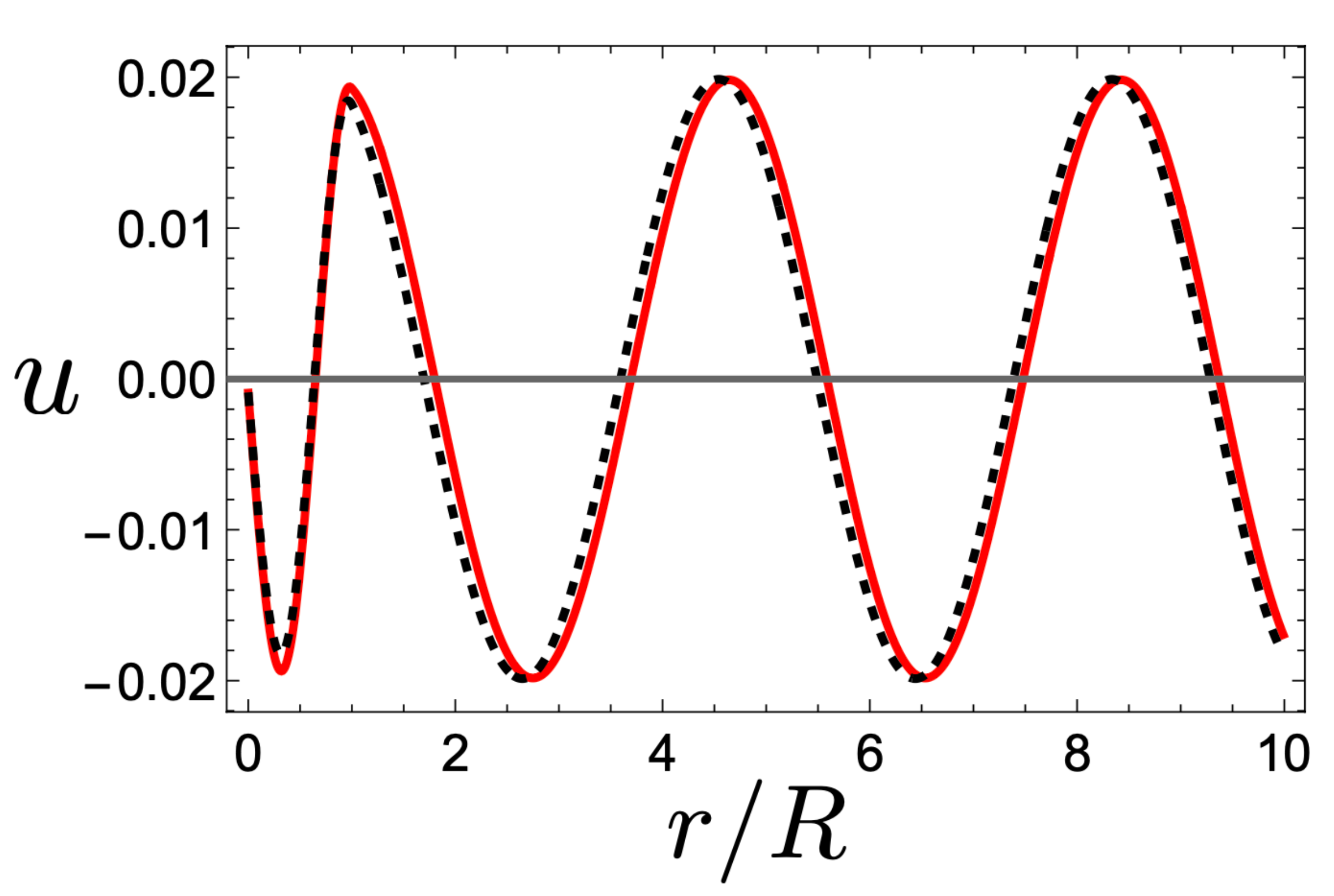}
\end{center}
\caption{(Color online) Comparison of scattering wavefunctions $u(r) = r \phi(r)$ of the open effective channel. In the continuous (red) line we show the numerical solution\cite{clarify} of the full effective open channel given by (\ref{Sch-id}) and in the dashed (black) line the analytical solution neglecting the integral term in the effective Hamiltonian (\ref{scatst}). We note that the main effect is a small phase difference in the region $r > R$. The calculation is for the same set of parameters used throughout the text and for $E = 1.4$. Units $\hbar = m= R=1$.
} \label{wavefcns}
\end{figure}

\begin{figure}[htbp]
\begin{center}
\includegraphics[scale=0.35]{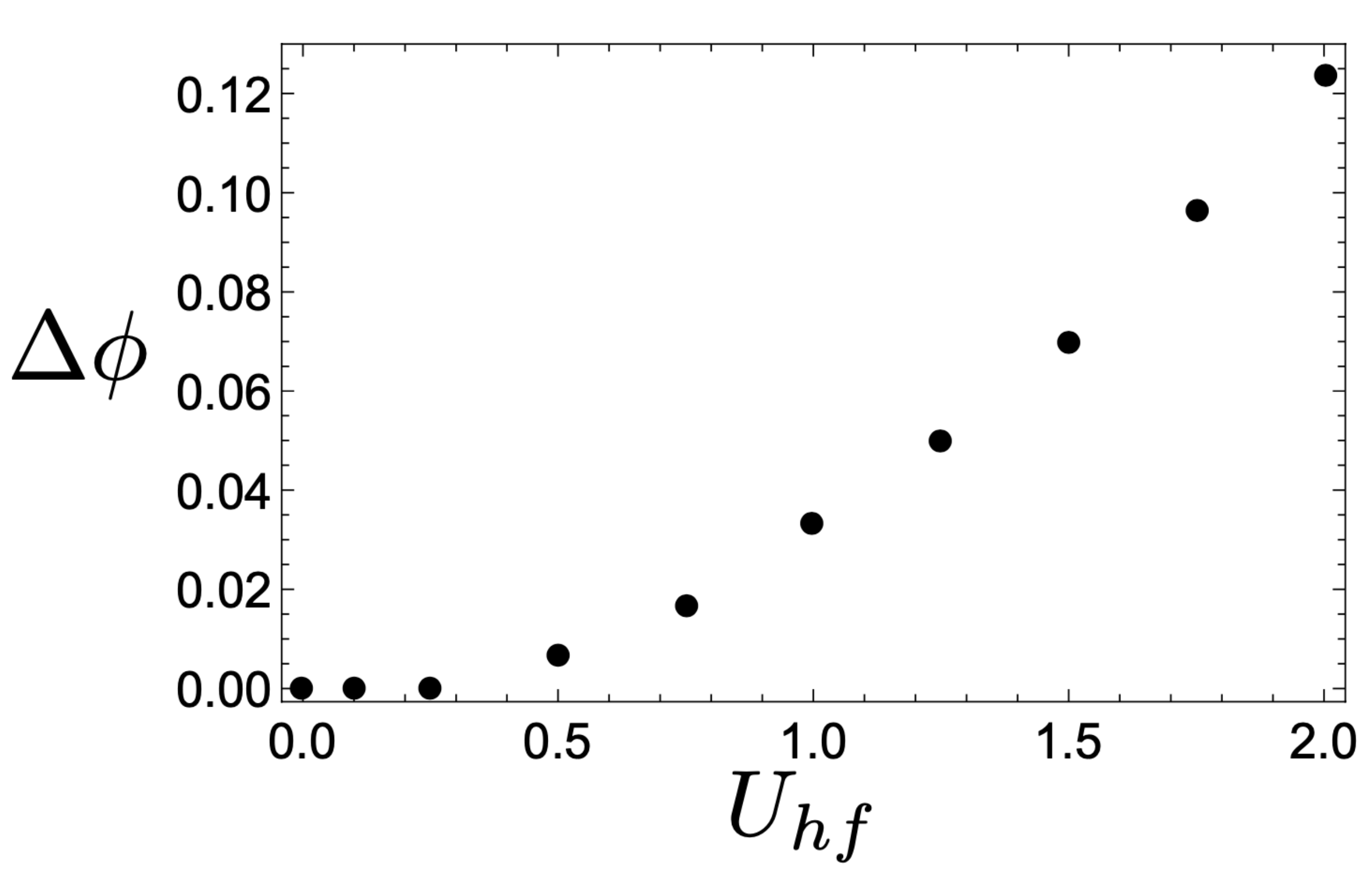}
\end{center}
\caption{(Color online) Differences of phases of numerical solutions\cite{clarify} of scattering wavefunctions of the effective open channel (\ref{Sch-id}), with and without the integral term in the effective Hamiltonian, as a function of the strength of the hyperfine interaction $U_{hf}$, for the same square-well potentials used throughout the text and for $E = 1.4$. See figure \ref{wavefcns}. Units $\hbar = m = R = 1$.
} \label{fases}
\end{figure}

Now we have to deal with the bound and scattering states, for $l = 0$, of the effective open channel, whose Hamiltonian is given by (\ref{Hefe0}). This gives rise to an integro-differential Schr\"odinger equation for $\psi(r) = \langle {\bf r} | \psi \rangle$
\begin{eqnarray}
\fl &&\left[-\frac{\hbar^2}{2m}\nabla^2  +  \hat V_o(r) \right]  \psi(r) +\nonumber \\
&&    \int d^3 r^\prime \int_{\Delta \mu B}^\infty  \frac{d \epsilon}{\epsilon_0 - \epsilon}\>  V_{hf}(r)  \phi(\epsilon,r)\phi^*(\epsilon,r^\prime) V_{hf}(r^\prime)  \psi(r^\prime) = E  \psi(r) \>,\label{Sch-id}
\end{eqnarray}
where $\epsilon_0$ is the eigenenergy of the bound state and  $|\phi(\epsilon)\rangle$  are the states of the continuum of  the displaced closed channel discussed above. The complete set of this Hamiltonian is needed for the calculation of the width $\Gamma$ (\ref{sigma0}) and the energy shift $\Delta \epsilon_0$ (\ref{BEpos}) of the $S$ matrix. This is, in principle, a very difficult task due to the integral term of (\ref{Sch-id}). In the absence of such a term the full solution is given by (\ref{boundst}) and (\ref{scatst}) simply replacing $U = U_c$. Fortunately, the weak interaction limit comes into our rescue and such a calculation can then be made. In Figs. \ref{wavefcns} and \ref{fases} we show a comparison of the numerical solution\cite{clarify} of the full equation (\ref{Sch-id}) with the analytical one of the single open channel having neglected the integral term, for the same values of the potentials $U_o = 10.0 \> \hbar^2/mR^2$, $U_c = 4.0 \> \hbar^2/mR^2$, used in the text. 
Figure \ref{wavefcns} shows scattering solutions of both cases, for a given value of the incident energy $E$, and one observes that the inclusion of the integral term results in a phase difference. This shift becomes larger as the hyperfine interaction $U_{hf}$ increases, as expected, and in figure \ref{fases} we show the this behavior. We found that the solutions are very close for values $U_{hf} \lesssim 0.25 \> \hbar^2/mR^2$ and the integral term in (\ref{Sch-id}) can be neglected. Nevertheless, in the analysis of the text we used $U_{hf} = 0.75 \> \hbar^2/mR^2$, a slightly larger value, and we found good agreement of the exact and the Feshbach expressions for the cross sections. It is important to stress that the usefulness of having made such an approximation allows for the calculation energy shift $\Delta \epsilon_0$ given in (\ref{Deltae0}), since this requires the knowledge of all the states of the open channel.

\newpage


\section*{References}


\begin{thebibliography}{99}

\bibitem{Ketterle} Davis K B, Mewes M-O,  Andrews M R,  van Druten N J, Durfee D S,  Kurn D M and  Ketterle W 1995 {\it Phys. Rev. Lett.} {\bf 75}

\bibitem{Cornell} Anderson M H, Ensher J R, Matthews M R, Wieman C E and Cornell, E A 1995 {\it Science} {\bf 269} 198 

\bibitem{Hulet} Bradley C C,  Sackett C A and Hulet, R G 1997 {\it Phys. Rev. Lett.} {\bf 78} 985 

\bibitem{Jin} Greiner M, Regal C A and  Jin D S 2003 {\it Nature} {\bf 426}

\bibitem{Kett-Zwier} Ketterle W and Zwierlein M W 2008 {\it Riv. Nuovo Cimento} {\bf 31} 247-422

\bibitem{Hulet-rev} Hulet R G, Nguyen J H V and  Senaratne R 2020 {\it Rev. Sci. Instrum.} {\bf 91}  011101

\bibitem{Pethick}   Pethick C J and Smith H 2008 {\it Bose–Einstein condensation in dilute gases} (Cambridge: Cambridge university press)


\bibitem{Dalfovo} Dalfovo F, Giorgini S, Pitaevskii L P and Stringari S 1999 {\it Rev.
Mod. Phys.} {\bf 71} 463

\bibitem{Bloch}  Bloch I, Dalibard J and  Zwerger W 2008  {\it Rev. Mod. Phys.} {\bf 80} 885 

\bibitem{Legget} Leggett A J 1980 in {\it Modern Trends in the Theory of Condensed Matter},  edited by Pekalski  A and Przystawa J A (Berlin: Springer) 13–27.

\bibitem{Giorgini} Giorgini S, Pitaevskii L P and Stringari S 2008 {\it Rev. Mod. Phys.} {\bf 80} 1215

\bibitem{Ohashi} Ohashi Y, Tajima H and van Wyk P 2020 {\it Prog. Part. Nucl. Phys.} {\bf 111} 103739

\bibitem{nobel} Lee D M, Osheroff D D and Richardson R C 1996 {\it Nobel Lectures} 

(https://www.nobelprize.org/uploads/2018/06/lee-lecture-2.pdf, 
https://www.nobelprize.org/uploads/2018/06/osheroff-lecture.pdf, 
https://www.nobelprize.org/uploads/2018/06/richardson-lecture-1.pdf)

\bibitem{Tiurev}  Tiurev K et al 2018 New J. Phys. {\bf 20} 055011

\bibitem{Muller} Bednorz J G and K. A. M\"uller K A 1986  {\it Z. Phys. B} {\bf 64} 189

\bibitem{Randeria} Randeria M, Duan J-M and Shieh L-Y 1990 {\it Phys. Rev. B} {\bf 4} 327

\bibitem{Tsuei} Tsuei C C and Kirtley J R 2000 {\it Rev. Mod. Phys.} {\bf 72} 969 

\bibitem{Chamel} Chamel N and Haensel P 2008 {\it Living Rev. Relativ.} {\bf 11} 1 

\bibitem{Strinati} Strinati G C, Pieri P, Roepke G, Schuck P and Urban M 2018  {\it Phys. Rep.} {\bf 738} 1


\bibitem{Verhaar} Tiesinga E, Verhaar B J, and Stoof H T C 1993 {\it Phys. Rev. A} {\bf 47} 4114

\bibitem{Moerdijk} Moerdijk A J, Verhaar B J, and Axelsson A 1995 {\it Phys. Rev. A} {\bf 51} 4852

\bibitem{Houbiers} Houbiers M, Stoof H T C, McAlexande W I, and Hulet R G 1998 {\it Phys. Rev. A} {\bf 57} R1497


\bibitem{OHara} O’Hara K M, Hemmer S L, Granade S R, Gehm M E, Thomas J E, Venturi V, Tiesinga E, and Williams C J 2002
{\it Phys. Rev. A} {\bf 66} 041401(R)

\bibitem{Lange} Lange A D, Pilch K, Prantner A, Ferlaino F, Engeser B, N\"agerl H-C, Grimm R and Chin C 2009
{\it Phys. Rev. A} {\bf 79} 013622 

\bibitem{Zurn} Z\"urn G, Lompe T, Wenz A N, Jochim S, Julienne P S and  Hutson J M 2013
{\it Phys. Rev. Lett.} {\bf 110} 135301

\bibitem{Julienne} Julienne  P S and Hutson J M 2014
{\it Phys. Rev. A} {\bf 89} 052715


\bibitem{Fesh1} Feshbach H 1958 {\it Ann. Phys.} {\bf 5} 357-390

\bibitem{Fesh2} Feshbach H 1962 {\it Ann. Phys.} {\bf 19} 287-313

\bibitem{Fesh3} Feshbach H 1967 {\it Ann. Phys.} {\bf 43} 410-420


\bibitem{Timmermans}  Timmermans E, Tommasin P, Hussein M and Kerman A 1999 {\it Phys. Rep.} {\bf 315} 199-230

\bibitem{Duine} Duine R A, Stoof H T C 2004 {\it Phys. Rep.} {\bf 396} 115-195

\bibitem{Chin} Chin C, Grimm R, Julienne P, and Eite Tiesinga E 2010 {\it Rev. Mod. Phys.} {\bf 82} 1225 


\bibitem{LLQM} Landau L D and Lifshitz E M 1981 {\it Quantum Mechanics (Non-relativistic theory)} (Oxford: Pergamon Press)

\bibitem{Liboff} Liboff  R L 1998 {\it Introductory Quantum Mechanics} (Oakland: Holden-Day)

\bibitem{Cohen} Cohen-Tannoudji C, Diu B and Lalo\"e F 1977 {\it Quantum Mechanics, Volume II} (New York: John Wiley \& Sons)

\bibitem{Newton} Newton R G 1982 {\it Scattering Theory of Waves and Particles} (New York: Springer-Verlag)

\bibitem{Taylor} Taylor J R 2000 {\it Scattering Theory. The Quantum Theory of Nonrelativistic Collisions} (Mineola: Dover Publications, Inc.)



\bibitem{Kokkelmans} Kokkelmans S J J M F, Milstein J N, Chiofalo M L, Walser R, and Holland M J 2002 {\it Phys. Rev. A} {\bf 65} 053617

\bibitem{Gurarie} Gurarie  V and Radzihovsky L 2007 {\it Ann. Phys.} 2–119

\bibitem{Wasak} Wasak T, Krych M, Idziaszek Z, Trippenbach M, Avishai Y and Band Y B
{\it Phys. Rev. A } {\bf 90} 052719 

\bibitem{Fraser} Fraser P A and Burley S K 1982 {\it Eur. J. Phys.} {\bf 3} 230

\bibitem{Taron} Taron J 2013 {\it Am. J. Phys.} {\bf 81} 603–609

\bibitem{Breit} Breit G and Wigner E 1936 {\it Phys. Rev.} {\bf 49} 519

\bibitem{delta} Use the identity $\lim_{\varepsilon \to 0} \frac{1}{x + i \varepsilon} = {\cal P} \frac{1}{x} - i \pi \delta(x)$.

\bibitem{clarify} The exact solution of the full coupled Hamiltonian (\ref{H}), with potentials (\ref{potentials}), can be numerically solved with {\it Mathematica}\cite{Math}. The solution of the effective Hamiltonian (\ref{Sch-id}), used to compute the wave function of figure (\ref{wavefcns}) and phase shifts of (\ref{fases}), does require a bit more of numerical work since such an equation is an integro-differential one; see \cite{Grover} for an example of a numerical method to solve (\ref{H}). In a lecture class it can be assumed that the integral part of the effective Hamitonian can be neglected, then all calculations can be performed with {\it Mathematica}.

\bibitem{Math} Wolfram Research, Inc. {\it Mathematica}, Version 13.0. Provided by UNAM.

\bibitem{Grover} Andrade-S\'anchez  G 2023 {\it Master Thesis} UNAM. 

\end{thebibliography}
\end{document}